\batchmode
\makeatletter
\def\input@path{{/Users/costin/Desktop/phd/year3/writing/cast/paper-named-casts/explicit_cast_paper_lyx/}}
\makeatother
\documentclass[twoside,english,british,twocolumn,3p]{elsarticle}
\usepackage[T1]{fontenc}
\usepackage[latin9]{inputenc}
\usepackage{babel}
\usepackage{array}
\usepackage{prettyref}
\usepackage{booktabs}
\usepackage{url}
\usepackage{multirow}
\usepackage{graphicx}
\usepackage{subscript}
\usepackage[pdftex,unicode=true]
 {hyperref}

\makeatletter

\providecommand{\tabularnewline}{\\}

\journal{Journal of Systems and Software}

\usepackage{booktabs}
\usepackage{subfig}
\AtBeginDocument{%
 \let\prettyref\autoref%
  \providecommand\BibTeX{{%
    \normalfont B\kern-0.5em{\scshape i\kern-0.25em b}\kern-0.8em\TeX}}}

\usepackage{pgfgantt}
\usepackage{tikz}

\usepackage{graphicx} % Required for including images
\graphicspath{{Figures/}} % Set the default folder for images

\usepackage{enumitem} % Required for manipulating the whitespace between and within lists

\usepackage{lipsum} % Used for inserting dummy 'Lorem ipsum' text into the template

\usepackage[linguistics]{forest}
\usepackage{amsmath,amsthm} % For including math equations, theorems, symbols, etc
\usepackage{multirow}
\usepackage{hyperref}
\usepackage{varioref} % More descriptive referencing
\usepackage{afterpage}
\usepackage{float}
\usepackage{listings}
\usepackage{xcolor}
\usepackage{url} 

\usepackage{blindtext}
\renewcommand\@biblabel[1]{}
\theoremstyle{definition} % Define theorem styles here based on the definition style (used for definitions and examples)

\theoremstyle{plain} % Define theorem styles here based on the plain style (used for theorems, lemmas, propositions)

\theoremstyle{remark} % Define theorem styles here based on the remark style (used for remarks and notes)

\hypersetup{
colorlinks=true, breaklinks=true, bookmarks=true,bookmarksnumbered,
urlcolor=webbrown, linkcolor=RoyalBlue, citecolor=webgreen, % Link colors
pdftitle={}, % PDF title
pdfauthor={\textcopyright}, % PDF Author
pdfsubject={}, % PDF Subject
pdfkeywords={}, % PDF Keywords
pdfcreator={pdfLaTeX}, % PDF Creator
pdfproducer={LaTeX with hyperref and ClassicThesis} % PDF producer
}

\definecolor{backcolour}{rgb}{0.95,0.95,0.92}
\definecolor{codegray}{rgb}{0.25,0.25,0.25}

\@ifundefined{showcaptionsetup}{}{%
 \PassOptionsToPackage{caption=false}{subfig}}
\usepackage{subfig}
\makeatother

\usepackage{listings}
\addto\captionsbritish{}
\addto\captionsenglish{}

\begin{document}

\begin{frontmatter}{}

\title{Do Names Echo Semantics? A Large-Scale Study of Identifiers\\ }

\title{Used in C++'s Named Casts.}

\author[affiliation1]{Constantin Cezar Petrescu\corref{cor1}}
\ead{c.petrescu@surrey.ac.uk}
\author[affiliation1]{Sam Smith}
\ead{ss02680@surrey.ac.uk}
\author[affiliation1]{Rafail Giavrimis}
\ead{r.giavrimis@surrey.ac.uk}
\author[affiliation2]{Santanu Kumar Dash}
\ead{santanu.dash@rhul.ac.uk}

\address[affiliation1]{University of Surrey, Guildford, Surrey, United Kingdom}
\address[affiliation2]{Royal Holloway, University of London, Egham, Surrey, United Kingdom}

\cortext[cor1]{Corresponding author}
\begin{abstract}
Developers relax restrictions on a type to reuse methods with other
types. While type casts are prevalent, in weakly typed languages such
as C++, they are also extremely permissive. Assignments where a source
expression is cast into a new type and assigned to a target variable
of the new type, can lead to software bugs if performed without care.
In this paper, we propose an information-theoretic approach to identify
poor implementations of explicit cast operations. Our approach measures
accord between the source expression and the target variable using
conditional entropy. We collect casts from 34 components of the Chromium
project, which collectively account for 27MLOC and random-uniformly
sample this dataset to create a manually labelled dataset of 271 casts.
Information-theoretic vetting of these 271 casts achieves a peak precision
of 81\% and a recall of 90\%. We additionally present the findings
of an in-depth investigation of notable explicit casts, two of which
were fixed in recent releases of the Chromium project.
\end{abstract}
\begin{keyword}
C++ Type Conversions\sep Languages\sep Program Analysis
\end{keyword}

\end{frontmatter}{}

\section{Introduction}

Developers like flexibility while using programming language features
during software development. Type casts allow developers to work around
the restrictions imposed on a specific type and use methods written
for other types. While casting offers flexibility, it can lead to
undefined behaviour in weakly typed languages like C/C++. For example,
considering the cast operation \lstinline*a=(T)b*, the outcome of
this statement is unclear unless we know what \lstinline*T* stands
for and what are the types of \lstinline*a* and \lstinline*b*. If
\lstinline*a* and \lstinline*b* are scalars, this could be a value
conversion. If they are objects, this could be a downcast from \lstinline+b+
to create \lstinline+a+, if \lstinline*a*'s class is derived from
\lstinline*b*'s class. \lstinline*a* and \lstinline*b* could be
unrelated pointer types, in which case, the set of permissible operations
is so vast that compilers might struggle to identify semantic errors. 

\citeauthor{typesafety_study} \citeyearpar{typesafety_study} studied
the safety of type casts and found that a quarter of them were guarded
with type checks to ensure their validity of type casts against run-time
errors. This was corroborated in a later study by \Citeauthor{casting_java_explicit}
\citeyearpar{casting_java_explicit} on the classification of patterns
for type casting. A study of implicit casting in JavaScript {[}\citeauthor{js_study}
\citeyear{js_study}{]} found most implicit casts to be harmless and
useful, implying that developers use them judiciously. \citeauthor{casting_java_explicit}
\citeyearpar{casting_java_explicit} performed a study of how developers
use type casts in Java and found 26 usage patterns for type casts.
Importantly, they discovered that half of the casts inspected by them
were not guarded locally which could potentially cause run-time errors.
Thus, there is a need to vet type casts to understand if they are
being used carefully. 

Type casts come in two forms: \emph{implicit} and \emph{explicit}.
Implicit casts or coercions are conversions from one type to another
without explicitly specifying the new type, and they are usually limited
to numeric types. Explicit casts are type conversions where syntactic
constructs are used to specify the target type. Compilers have multiple
checks to vet implicit casts on numerics. Even so, it is not possible
to categorically enforce checks on casts for several mainstream languages
with user-defined types. Therefore, for languages like C++, that are
permissive in how memory is used at a low-level, several primitives
for explicit type conversion have been introduced. These primitives,
which are called \emph{named casts}, come with a unique set of checks
on the cast operation. They are the recommended technique for explicitly
changing one type to another in C++ and have two placeholders in the
primitive: a \emph{source} expression that needs to be cast and the
\emph{destination} type for the cast. 

In this paper, we propose a lightweight approach to check if casts
are used judiciously. \citeauthor{dual_channel} \citeyearpar{dual_channel}
presented source code as being dual channel. One channel is represented
by the algorithmic channel comprised of instructions understood and
executed by computers. The second channel is the natural language
channel which consists of identifiers and comments to provide semantics
for the instructions. In line with the recent work that uses meaning
in identifiers in programs {[}\citeauthor{refinym} \citeyear{refinym};
\citeauthor{flexeme} \citeyear{flexeme}{]}, we propose a dual channel
approach to analyse \emph{named casts}. Our assumption is that developers
leave hints about their intent in the identifiers that they choose
and that this information can be used to check fidelity of an explicit
type conversion. In particular, we are interested in knowing if the
\emph{source} expression that is being cast is related to the \emph{destination}
variable to which the result of the cast is being assigned. Our main
contributions are as follows: 
\begin{enumerate}
\item We propose an information-theoretic approach and a tool to detect
poor programming and naming practices in named casts. Our tool's code,
the dataset we evaluated it on and the results of the information-theoretic
analysis are available online \footnote{\url{https://github.com/Constantin-Petrescu/CastChecker/}}.
\item We extract named casts from the Chromium project {[}\citeauthor{chrome}
\citeyear{chrome}{]}, which is an aggregation of over 34 components
with nearly 27 million lines of C++ code. We sample this dataset random-uniformly
at 90\% confidence (5\% margin of error) and create a dataset of 271
named casts vetted by human developers to create a ground truth. 
\item We propose a taxonomy of how a user can build a named cast operation
along with a set of broad categories of cast usage. We show how the
taxonomy fits the ground truth set to provide insights on how casts
are used. 
\item We evaluate our tool on the vetted dataset and achieve a peak accuracy
of 92\%. Additionally, we perform an in-depth investigation of the
cases flagged by the tool and provide insights into both poor programming
practices and poor naming choices when using named casts.
\item We demonstrate impact through two instances of named casts flagged
by the tool that have been patched in a recent release of the software.
In addition, we discover another two cases where the named casts were
part of code with a high complexity that eventually led to bugs. After
the bugs were fixed, the named casts were completely removed. 
\end{enumerate}
We discuss an overview of casting in C++, along with an example of
imprecise \emph{named cast} usage and the motivation for our research
in \hyperref[subsec:Context]{Section 2}. We describe our methodology
in  \hyperref[subsec:Methodology]{Section 3} and the results of
our evaluation in \hyperref[subsec:Evaluation]{Section 4}. \hyperref[subsec:Limitations]{Section 5}
discusses some threats to validity. \hyperref[subsec:Related-Work]{Section 6}
presents the related work and \hyperref[subsec:Conclusion]{Section 7}
concludes this work.

\section{Cast operations, their use and the motivation of the work\label{subsec:Context}}

C++ provides several ways in which a type conversion can be effected.
We first provide an overview of these ways. Then, we show through
an example how, despite clear guidelines on how casts should be used,
type casts can be used imprecisely.

\subsection{Implicit and Explicit Casts\label{subsec:Type-conversions}}

Type conversions are operations where the type of an expression is
changed from one type to another. There are two types of conversions:
implicit and explicit casts. In implicit casts, the conversion is
done without the developers explicitly specifying the type to which
a value needs to be converted. Implicit casts are performed automatically
by the compiler if there is a viable conversion. For example, in C/C++,
it is possible to pass a \lstinline*float* as an argument to a method
which expects a \lstinline*double* {[}\citeauthor{implicitCastCPP}
\citeyear{implicitCastCPP}{]}. Implicit conversions, also known as
standard conversions {[}\citeauthor{implicitCastSTD} \citeyear{implicitCastSTD}{]},
are generally applied on built-in numerical data types, booleans and
some pointer conversions {[}\citeauthor{implicitCastDocumentation}
\citeyear{implicitCastDocumentation}{]}. The implicit conversions
between numerical types are called promotions {[}\citeauthor{implicitCastSTD}
\citeyear{implicitCastSTD}{]} and are allowed from smaller size types
to larger size types. 

C/C++ also allows explicit conversion using syntactic constructs.
The syntactic constructs tell the compiler to perform a type conversion
where the new type is specified in comparison to implicit conversions.
There are two ways to perform explicit casts, which are presented
in \prettyref{fig:syntaxCasting}. Here, a variable \lstinline*x*
of type \lstinline*double* is converted to an \lstinline*int* type.
The first is the functional style, where the target type is treated
as a method and the variable that will be converted is passed as an
argument. The other is commonly referred to as the C-style syntax
where the use of the variable is qualified by the target type within
parenthesis. 

\begin{lstlisting}%
[caption={Functional and C-style syntax for implicit type conversion.},float=t,label={fig:syntaxCasting}]
double x = 10.3; 
int y;
y = int (x);    // functional notation 
y = (int) x;    // c-like cast notation   
\end{lstlisting}

\begin{lstlisting}%
[caption={Example of \lstinline*static_cast* .},float=t,label={fig:staticCast}]
class Base {}; 
class Derived: public Base {};
Base * a = new Base; 
Derived * b = static_cast<Derived*>(a);         
\end{lstlisting}

The function and C-style explicit casts can handle conversion of built-in
types in the compiler such as numeric types. However, using those
operators on user-defined types, and in particular, class hierarchies,
requires additional language constructs, and named cast operators
were introduced in C++. There are four named cast operators in C++:
\lstinline*static_cast*, \lstinline*dynamic_cast*, \lstinline*const_cast*
and \lstinline*reinterpret_cast*. Out of the four, \lstinline*static_cast*,
\lstinline*dynamic_cast* and \lstinline*const_cast* perform additional
checks either statically or at runtime to avoid undefined behaviour,
because of incorrect usage of type casts {[}\citeauthor{explicitCastCPP}
\citeyear{explicitCastCPP}{]}. \lstinline*reinterpret_cast* is the
most permissive with no checks on the validity of the type conversion.
It merely reinterprets the memory holding an object as another type.

\paragraph*{The \lstinline*static_cast* operator}

\lstinline*static_cast* vets the casts by statically checking the
validity of the conversions against the class hierarchies {[}\citeauthor{staticCastCPP}
\citeyear{staticCastCPP}{]}. As shown in \prettyref{fig:staticCast},
a downcast of an object \lstinline*a* typed as base class \lstinline*Base*
to a derived class \lstinline*Derived* is allowed, but the developer
needs to be confident that \lstinline*a* will never be an object
of another derived class of \lstinline*Base*. If the latter happens,
accessing a field of the \lstinline*Derived* class through \lstinline*b*
would lead to undefined behaviour. This is because \lstinline*static_cast*
does not apply runtime checks to validate if \lstinline*a* is an
object of type \lstinline*Derived* or another derived class \lstinline*Derived2*
of \lstinline*Base*. Therefore, the correctness of a \lstinline*static_cast*
is reliant on the developer. \lstinline*static_cast* operations are
also used for converting \lstinline*enum* and \lstinline*void* types
where the developer is sure of the type of the data pointed to by
a \lstinline*void* pointer. 

\begin{lstlisting}%
[caption={Example of \lstinline*dynamic_cast*.},float=t,label={fig:dynamicCast}]
class Base { virtual void vf(){} };

class Derived : public Base { };
int main()  
{   
	Base *pBDerived = new Derived;   
	Derived *pd;
	pd = dynamic_cast<Derived*>(pBDerived);
	return 0; 
}
\end{lstlisting}

\paragraph*{The \lstinline*dynamic_cast* operator}

\lstinline*dynamic_cast* is an operator used for casting pointers
and class reference conversions. Unlike \lstinline*static_cast*,
\lstinline*dynamic_cast* checks whether the \emph{named cast} is
permissible at runtime. If not, it returns a null pointer {[}\citeauthor{dynamicCastCPP}
\citeyear{dynamicCastCPP}{]}. This operation guarantees that the
result points to a valid object of the new type at the end of the
type conversion. \prettyref{fig:dynamicCast} presents an example
of \lstinline*dynamic_cast* for a pointer \lstinline*pBDerived*.
The pointer has the initial type \lstinline+Base*+ and it points
to a \lstinline*Derived* object. Through the cast on Line 8, the
\lstinline*pBDerived* pointer becomes an object of class \lstinline*Derived*.
\lstinline*dynamic_cast* operations perform validity checks using
the Run-Time Type Identification (RTTI) which is a feature in C++
to inspect types of objects at runtime. Naturally, the runtime checks
introduce overheads and \lstinline*dynamic_cast* is an expensive
operation for performance-sensitive applications. 

\paragraph*{The \lstinline*reinterpret_cast* operator}

This operator's role is to reinterpret the memory holding an object
of one type as another type to convert it from one type to another.
The pointer to the memory is recast into a new pointer type without
any checks if the content can be of the new type. In general, this
cast is used on low-level conversions based on a reinterpretation
of the binary values of the variables {[}\citeauthor{reinterpretCastCPP}
\citeyear{reinterpretCastCPP}{]}. In \prettyref{fig:reinterpretCast},
a \lstinline*reinterpret_cast* example is shown on Line 5. The variable
\lstinline*a* of class \lstinline*A* is reinterpreted to type \lstinline*B*
and assigned to pointer \lstinline*b* even though \lstinline*A*
and \lstinline*B* are unrelated in the class hierarchy. The \lstinline*reinterpret_cast*
has a lower overhead than the other operators since it does not perform
validity checks. Like the \lstinline*static_cast*, though, the correctness
for this conversion relies entirely on the developer.

\begin{lstlisting}%
[caption={Example of \lstinline*reinterpret_cast*.},float=t,label={fig:reinterpretCast}]
class A { /* ... */ };
class B { /* ... */ };

A * a = new A;
B * b = reinterpret_cast<B*>(a);
\end{lstlisting}

\paragraph*{The \lstinline*const_cast* operator}

This operator makes it possible to modify variables that have the
type qualifier \lstinline*const*, which directs the compiler not
to allow any modification for a variable, and \lstinline*volatile*,
which prevents the compiler from applying any optimisations on the
variable. An example is presented in \prettyref{fig:constCast}. The
variable \lstinline*c* of type \lstinline*const char** is passed
as an argument to a method \lstinline*print* which only supports
\lstinline+char*+. This forces the use of \lstinline*const_cast*
in Line 9 as mandatory to match the actual type to the formal parameter
type. The C++ standard states that the \lstinline*const_cast* operator
can introduce undefined behaviour in programs. This situation can
appear if the constness is removed from a variable and after the variable
is modified {[}\citeauthor{constCastCPP} \citeyear{constCastCPP}{]}. 

\begin{lstlisting}%
[caption={Example of \lstinline*const_cast*.},float=t,label={fig:constCast}]
void print (char * str)
{
  cout << str << '\n';
}

int main () 
{
  const char * c = "sample text";
  print ( const_cast<char *> (c) );
  return 0;
}
\end{lstlisting}

\subsection{An example of imprecise named cast usage\label{subsec:An-example-of}}

\begin{lstlisting}%
[caption={An example where two \lstinline*static_cast* operators are used to
iterate over an enumeration and store integer values in a dictionary.
The snippet is from the file \emph{net\_log\_util.cc} of component
\emph{Net} taken from an open source implementation of the QUIC protocol
in the Chromium project},float=t,label={fig:motiv}]
// Add information on the relationship between QUIC error codes
// and their symbolic names.
std::unique_ptr<base::DictionaryValue> dict(new base::DictionaryValue());

for (QuicErrorCode error = QUIC_NO_ERROR; 
      error < QUIC_LAST_ERROR;
      error = static_cast<QuicErrorCode>(error + 1)) {
  dict->SetInteger(QuicErrorCodeToString(error), 
      static_cast<int>(error));
}
\end{lstlisting}

\emph{Named casts} were proposed initially to provide semantic clarity.
However, developers sometimes use them to bypass type system restrictions
at the cost of increased code complexity. Consider \prettyref{fig:motiv}
as an example. The code is a snippet taken from the implementation
of QUIC protocol {[}\citeauthor{quic} \citeyear{quic}{]}. QUIC is
a general-purpose transport layer network protocol open sourced as
a part of the Chromium project. There are two uses of the \lstinline*static_cast*
operator in this snippet, which populates a dictionary \lstinline*dict*
with key-value pairs, which are strings representing an \lstinline+error+
description and an integer representing the \lstinline+error+ code.
It is important to note here that \lstinline+error+ itself is neither
an integer nor a string but an \emph{unscoped} \lstinline*enum* type
\lstinline+QuicErrorCode+.

The type \lstinline*enum* or enumeration is a user-defined type which
consists of a set of named integral constants {[}\citeauthor{enumerationsCPP}
\citeyear{enumerationsCPP}{]}. Enumerations are generally used in
three situations: a single choice where the developer filters through
the choices with a switch statement, a multiple choice through C-style
bitsets, or as a type definition for integral types. In \prettyref{fig:motiv},
the type \lstinline*enum* is not used for any of the three situations,
but it is used to iterate over the enumeration values and populate
\lstinline*dict*. By design, C++ does not encourage the iteration
over objects of type \lstinline*enum* since it does not provide an
iterator. In the example, the iteration is achieved by implicitly
casting the loop control variable \lstinline+error+ into an integer,
incrementing it and casting it back to \lstinline+QuicErrorCode+
using a \lstinline*static_cast* in Line 7. In the loop expression,
\lstinline+QUIC_NO_ERROR+ and \lstinline+QUIC_LAST_ERROR+ are the
first and last elements of the enumeration. The second \lstinline*static_cast*
in Line 9 converts the variable \lstinline+error+ of type \lstinline+QuicErrorCode+
to an \lstinline+int+. It is used as a parameter for the function
\lstinline+SetInteger+, which populates the dictionary \lstinline+dict+
with key-value pairs. This is the second time that the developers
chose to cross the boundaries between an \lstinline*enum* type to
an \lstinline*int* to be able to use operators of the type \lstinline*int*. 

The iteration on \lstinline*enum* objects can be pernicious, as \lstinline*enum*
types are not guaranteed to be contiguous. The Clang++ compiler would
replace \lstinline+QUIC_NO_ERROR+ and \lstinline+QUIC_LAST_ERROR+
to their actual values in the loop from the snippet. This means that
\lstinline+error+ would take all the values in the corresponding
range. The enumeration \lstinline+QuicErrorCode+ is not contiguous
and the values for each entry are defined by the developers. This
means that the \lstinline+dict+ could contain error codes that were
not described originally in \lstinline+QuicErrorCode+. However, the
developers handle those cases explicitly in the function \lstinline+QuicErrorCodeToString+,
which contains a \lstinline*switch* over all the values from \lstinline+QuicErrorCode+.
This function returns the string of the \lstinline+error+ or an invalid
error code for any other values. This implementation is not erroneous;
however, it is suboptimal. 

One may wonder at this stage, what could be a better solution and
what should the solution aim to achieve? Type systems came about to
ensure type safety and casts typically should be avoided wherever
possible. The aim of a better solution should be to keep the \lstinline*enum*
and \lstinline*int* types separate and implement all operators essential
to iterate or operate in the \lstinline*enum* space. The developers
used an enumeration to generate a dictionary object type which will
be used later by the rest of the application. The enumeration implementation
consists of the \lstinline+QuicErrorCode+ declaration along with
a set of functions of switch cases such as \lstinline+QuicErrorCodeToString+
that allow the return of the string for an error. We believe a better
solution would be to declare and use a dictionary from the start rather
than declaring and using the enumeration to create the dictionary. 

This solution would not require the crossing of type boundaries, since
the type of the dictionary can be declared accordingly to the types
of the values. Also, the solution would bring improved efficiency.
Enumerations are efficient since they are resolved at  compile time
and converted into integral literals at the bitcode level. The enumerations
are used along with switch cases and iterations over the enumerations,
which present a linear efficiency. This efficiency performs well on
a small number of cases, which is not the case for \lstinline+QuicErrorCode+
since it consists of 199 cases. On the other hand, the selection of
a key in a dictionary would have a logarithmic efficiency. We are
not sure if \lstinline+QuicErrorCode+ is used in any other part of
the application, but dictionaries should generally perform better
than large enumerations. Our solution would also ease the code maintainability
process. Each time \lstinline+QuicErrorCode+ needs to be updated,
it requires modifications at the declaration and at each function
with switch cases. It would be easier to maintain a dictionary since
the only modification required would be at the declaration. This example
shows a need for tools that identify if the cast of types is essential
and if the cast is done correctly. It is crucial to ensure that the
crossing type boundaries are beneficial from a software engineering
point of view, allowing code reuse without confounding the uses of
types and operators for those types.

\subsection{Motivation\label{subsec:Motivation}}

\begin{figure*}[!t]
\includegraphics[width=1\textwidth]{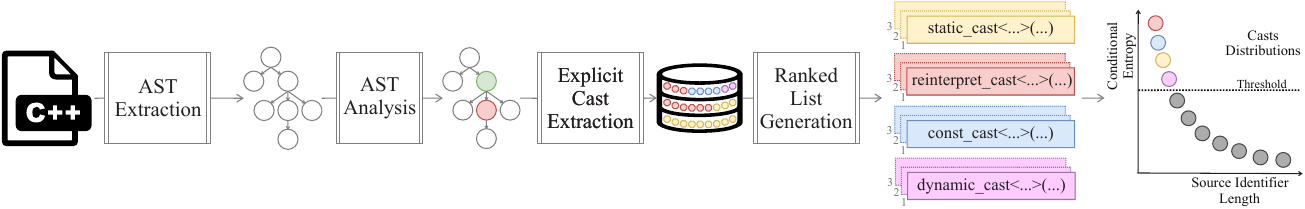}\caption{Software architecture diagram of our tool which extracts named casts
from a C++ codebase and analyses them using information theory.}
\label{fig:arh_diagram}
\end{figure*}

In this research, we hypothesise that in large and mature projects
such as Chromium, where code is reviewed before it is merged in the
application, there are hints in program identifiers that point to
their purpose. We aim to use this natural language information in
identifiers to understand if \emph{named casts} are being used for
good software engineering reasons. If this is not the case then we
aim to identify when the named casts are being poorly implemented,
such as when they are part of unoptimised or buggy code. For example,
the actual to formal binding for the method \lstinline+SetInteger+
binds \lstinline+error+ of type \lstinline+QuicErrorCode+ to a formal
named \lstinline+in_value+ of type \lstinline*int*. A perfunctory
check of the names for the variables and the types may seem that these
variables are disparate. However, one may notice upon close inspection
that \lstinline+SetInteger+ is a modifier of a dictionary. Therefore,
it is essential that formal arguments of this modifier are named generically.
In this work, we combine an automated analyser with human inspection
to classify cases where \emph{named casts} are used to point out both
good and poor practices in using \emph{named casts}.

In a \emph{named cast} situation, precise names are meaningful names
that reflect the relation between the \emph{source} and \emph{destination}.
The choice of the identifiers is not only vital during development,
but also during maintenance. Precise names reflect that the developers
had a good understanding of the problem that they solved. The same
precise names allow other developers to gain a faster and more comprehensive
understanding of the code. Thus, the reusability and maintenance of
the code is made easier. If the relation between \emph{source} and
\emph{destination} does not exist, developers may be misled by the
names and overlook some cases which could be dangerous during code
testing and maintenance. For instance, consider the following set
of pairs of imprecise identifier names: \lstinline*tag* with \lstinline*chars[i]*,
\lstinline*levels* with \lstinline*fparams[0]*, \lstinline*param*
with \lstinline*bufSize*, \lstinline*t* with \lstinline*output_cursor*,
\lstinline*val* with \lstinline*p[i]*, \lstinline*frames* with
\lstinline*out_trace*. All of these pairs exhibit no meaningful connection
or semantic similarity, and this could potentially lead to misunderstanding.
Those cases need to be identified and refactored with meaningful names.
Our tool uses the information-theoretic analysis to discover imprecise
names given the \emph{source} expression and \emph{destination} variable.

\section{Methodology\label{subsec:Methodology}}

Our objective is to analyse if natural language identifiers are indicative
of the purpose of the cast. For this, we focus on assignment expressions
where the right hand side is a named cast expression and on actual-to-formal
bindings in method calls where the argument to the method is a named
cast expression. In both cases, the expression that is cast to a new
type is referred to as the \emph{source} and the identifier to which
the cast expression is bound is called the \emph{destination}. 

\subsection{Software Architecture of the Proposed Tool}

\prettyref{fig:arh_diagram} presents an overview of our tool. We
rely on a Clang plugin to traverse the abstract syntax tree (AST)
of source files. Our plugin traverses every node to discover named
cast expressions and then determines if the expression is part of
a larger sub-tree representing an assignment operation or a method
call expression. Details of this process can be found in \hyperref[subsec:Extraction-of-Named-Casts]{Section 3.2}.
We detail the analysis of the extracted dataset in \hyperref[subsec:Data-analysis]{Section 3.3}
and the results of our manual investigation can be found in \hyperref[subsec:Qualitative-analysis]{Section 4.5}.

Our corpus is generated from the Chromium project {[}\citeauthor{chrome}
\citeyear{chrome}{]}. Chromium is an extensive system written in
C++ and it only supports the Clang compiler for building. Chromium
uses the Ninja build system and GN {[}\citeauthor{gn} \citeyear{gn}{]}
as a meta-build system that generates Ninja build files. The Ninja
files run the Clang compiler, for which our analysis plugin is written,
on the C++ files. Therefore, we modified the meta-build system to
use a local version of Clang that is compatible with our plugin. The
output generated by our modified compilation phase is a JSON file
containing the named cast information for every C++ file that is compiled.
These named casts constitute the dataset for our analysis which is
described next.

\subsection{Extraction of Named Casts\label{subsec:Extraction-of-Named-Casts}}

\begin{figure}[t]
\begin{centering}
\includegraphics[scale=0.6]{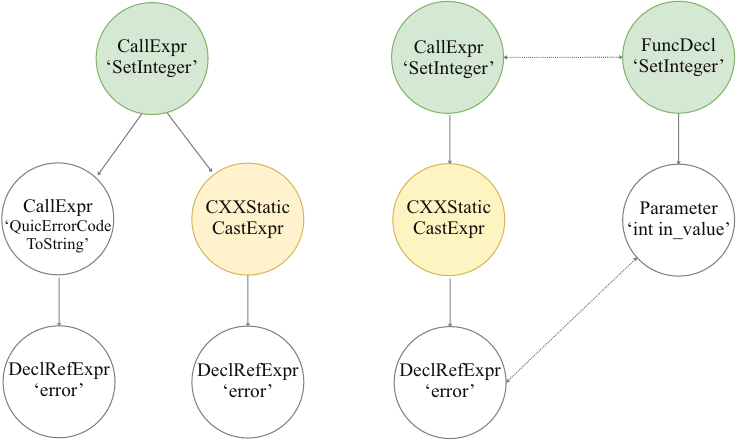}
\par\end{centering}
\caption{Abstract syntax tree representation for our motivating example; we
selected only the nodes of interest. The left side shows the function
call, \emph{SetInteger}. The right side presents the mapping between
the function call and the function definition.}

\label{fig:ast_diagram}
\end{figure}

In \prettyref{fig:ast_diagram}, we present an example of how our
plugin analyses a named cast from the \emph{Net}\textbf{\emph{ }}sub-system
in Chromium. After Clang parses the source file and produces an AST
for the file \emph{net\_log\_util.cc}, the plugin traverses the tree
and searches for named casts that are a part of either assignments
or call expressions. On the left in \prettyref{fig:ast_diagram},
the syntax tree for the function call \lstinline*SetInteger* is shown.
The node \lstinline*CallExpr* has a child \lstinline*CXXStaticCastExpr*
which represents the node for \lstinline*static_cast* implying that
the named cast is used as an argument for a function call. The plugin
then follows the call to find the method definition. A projection
of the AST for the method definition is shown on the right in \prettyref{fig:ast_diagram}.
The plugin then links the formal parameter to the actual parameter
for \lstinline*SetInteger* and discovers that the \emph{source} variable
is \lstinline*error* and the \emph{destination} variable is \lstinline*in_value*.
All the macro names in the code will be replaced with actual code
at the compilation stage {[}\citeauthor{preprocessor} \citeyear{preprocessor}{]}.
However, the physical location of the named casts would still point
to the macro's call. To solve this, our plugin is designed to follow
macro definitions, post their expansion, to discover named casts inside
macro definitions as well. For each C++ file analysed, the Clang plugin
generates a JSON file with information about named casts. Each JSON
entry in the file consists of the type of named cast i.e. \lstinline*static_cast*,
\lstinline*dynamic_cast* \lstinline*reinterpret_cast* or \lstinline*const_cast*.
It additionally contains the type and the subtokens for the \emph{source}
and the \emph{destination }expression. To generate the subtokens,
we extract all tokens from each expression and we preserve only identifiers,
keywords and literals tokens. Those tokens are split in subtokens
based on the camelcase and snakecase separators.

\subsection{Data Analysis\label{subsec:Data-analysis}}

In this research, we study if the identifiers convey the reason for
the use of a named cast. We do this by comparing the \emph{source}
expression subtokens with the \emph{destination} variable subtokens.
Our comparison is based on a notion of entropy -- the amount of information
in names. We find cases where \emph{source }subtokens are significantly
different from the \emph{destination} subtokens. The difference is
measured using conditional entropy which computes the number of additional
bits that would be required to represent the \emph{destination} given
the subtokens in the \emph{source.} While we have access to the type
information, we do not use this information in the calculation of
the conditional entropy. The reason for this is that, during development
and sometimes in static time, the type of a variable is not always
visible to the human. That is why including the type in our analysis
would make it different than the way a human would view code.

Next, we show how we compute the conditional entropy of \lstinline+fooBar+
given the entropy for \lstinline+bazGoo+ in the named cast \lstinline+fooBar = static_cast<Quux*> bazGoo+.
\prettyref{eq:ent} presents the standard Shannon's formula for computing
the entropy {[}\citeauthor{shannonEntropy} \citeyear{shannonEntropy}{]},
which is the negative sum of the probabilities multiplied with the
logarithm value of the probability. Here, $X$ represents \lstinline+bazGoo+
and $x_{i}$ represents the probabilities for \lstinline+baz+ and
\lstinline+Goo+ which are the subtokens of the identifier. The subtokens'
probabilities have a value of $\frac{1}{2}$ since there are only
two possible options. Thus, $H(bazGoo)=-(2*\frac{1}{2}*log_{2}\frac{1}{2})=-[1*(-1)]=1$.
In other words, we need only one bit to represent the two possible
options for the \emph{source} subtokens. 

\begin{table}
\begin{tabular}{|c|c|c|c|c|}
\hline 
S &
D &
H(S) &
H(S, D) &
H(D|S)\tabularnewline
\hline 
\hline 
baz &
foo &
0 &
1 &
1\tabularnewline
\hline 
baz &
fooBar &
0 &
1.58 &
1.56\tabularnewline
\hline 
baz &
fooBaz &
0 &
0.92 &
-0.08\tabularnewline
\hline 
bazGoo &
foo &
1 &
1.58 &
0.58\tabularnewline
\hline 
bazGoo &
goo &
1 &
0.92 &
-0.08\tabularnewline
\hline 
bazGoo &
bazGoo &
1 &
1 &
0\tabularnewline
\hline 
bazGoo &
fooBar &
1 &
2 &
1\tabularnewline
\hline 
bazGooFoo &
fooBar &
1.58 &
1.92 &
0.34\tabularnewline
\hline 
\end{tabular}

\caption{Examples for Conditional Entropy Calculations; \protect \\
S - \emph{Source}, D - \emph{Destination}, H - Entropy.}
\label{tbl:entropy-examples}
\end{table}

We then compute the conditional entropy as shown in \prettyref{eq:cond_ent}
{[}\citeauthor{condEnt_ent} \citeyear{condEnt_ent}{]}. The conditional
entropy is the amount of information (in bits) required to express
the outcome of a random variable knowing the outcome of another random
variable. In \prettyref{eq:cond_ent}, \lstinline*Y* is a placeholder
for the subtokens from \lstinline+foo+ and \lstinline+Bar+ in our
example. We try to compute the conditional entropy of \lstinline*Y*
given \lstinline*X* based on the chain rule. Thus, the conditional
entropy value is the entropy value of the \emph{source}'s subtokens
subtracted from the joint entropy value of both \emph{source} and
\emph{destination} subtokens. In current example, the joint entropy
is computed for all the subtokens \lstinline+baz+, \lstinline+Goo+,
\lstinline+foo+ and \lstinline+Bar+. $H(bazGoo,fooBar)=-(4*\frac{1}{4}*log_{2}\frac{1}{4})=-[1*(-2)]=2$.
The conditional entropy tells how many more bits are needed to represent
the additional subtokens that the \emph{destination} identifiers bring
knowing the \emph{source}'s subtokens. In the example, the conditional
entropy equals with the difference between the joint entropy and entropy
of the \lstinline+bazGoo+ and it has value one. Thus, the \emph{destination}
\lstinline+fooBar+ identifier will require an additional bit in order
to represent the two new additional subtokens. Examples of entropy
values for identifiers can be found in \prettyref{tbl:entropy-examples}. 

\begin{equation}
H(X)=-\sum\limits _{i=1}^{n}P(x_{i})*\log P(x_{i})\label{eq:ent}
\end{equation}

\begin{equation}
H(Y|X)=H(X,Y)-H(X)\label{eq:cond_ent}
\end{equation}

The role of conditional entropy value is to discover how different
a \emph{destination} expression is, compared to the \emph{source}
expression used in a named cast. Therefore, we compare the subtokens
of the \emph{destination} expression with the subtokens of the \emph{source}
expression for each named cast operation we collected from Chromium.
If we were to consider the subtokens across multiple named cast cases
in the conditional entropy calculation for each case, then the result
would not be the difference between \emph{source} and \emph{destination}.
The comparison would instead identify if the \emph{destination} expression
contains unique subtokens compared to \emph{source} subtokens from
all the cases. The chances that some of the \emph{destination} subtokens
appear in the subtokens from \emph{source} expression increases with
the addition of multiple \emph{source} expressions in the calculation
of the conditional entropy. 

The conditional entropy values of the \emph{destination} given the
\emph{source} enables the identification of cases where the \emph{source}
looks significantly different from the \emph{destination}. A low conditional
entropy value implies that \emph{source} and \emph{destination} subtokens
are similar. On the other hand, a high conditional entropy value means
they have few subtokens in common. If identifiers are used for different
purposes, under the assumption that names are chosen carefully, their
subtokens will also be different. We are interested in the cases where
the conditional entropy is high. Those cases should generally point
to clear instances where disparate names are used in the \emph{source
}and the \emph{destination} expressions. This is indicative of the
\emph{destination }variable serving a different purpose than the \emph{source
}expression.

One may wonder why we did not use a simpler distance metric such as
Levenshtein Distance (LD) instead of conditional entropy. LD uses
three operations: insertion, deletion, substitution and the edit distance
is the number of operations used to transform the input string into
the output string. It is sensitive to the ordering of subtokens. Subtoken
ordering is not important to us as we want only to check if the subtokens
are being reused from the \emph{source }in the \emph{destination}.
Whether an identifier is called \lstinline+thrown_type+ or \lstinline+type_thrown+
is immaterial to us, but it affects the Levenshtein distance.

\subsection{Flagging Casts for Inspection \label{subsec:Decision_factors}}

Our tool flags casts for inspection if the conditional entropy of
a \emph{destination} variable, given the \emph{source} expression,
is above a pre-specified threshold. We refer to these casts as flagged
casts. The tool is parametric in the threshold: a user can tune the
threshold to their needs. For our experiments, we optimise the threshold
based on the precision-recall curve, as detailed in \hyperref[subsec:pr-curve]{Section 4.3}. 

\subsubsection{Shortlisting Casts}

We apply a set of preprocessing filters before computing the conditional
entropy for a cast. These filters are essential to reduce the noise
in entropy calculations, which may artificially inflate conditional
entropy values and introduce false positives. Our filters are described
below. 

\paragraph{Casts from Literals}

The universe of literals is large and literals are likely to differ
from well named identifiers that they are being cast into. This increases
the conditional entropy values of the \emph{destination} given the
\emph{source}, introducing false positives. We check if the \emph{source}
is a literal and if it is, the cast is not flagged by the tool. The
possible literals in C++ are: integers in base 2, 10 or 16 (e.g. \lstinline*42*,
\lstinline*033*, \lstinline*0x00*), floating point numeric values
(e.g. \lstinline+0.5+, \lstinline*-0.26E-5*), characters (e.g. \lstinline+'a'+,
\lstinline*'\{'*), escape sequences (e.g. \lstinline*\\n*, \lstinline*\\r*),
and strings (e.g. \lstinline*"Hello World"*).

\paragraph*{Descriptive Variables and Types}

We check if the \emph{source} is part of the \emph{destination} using
the substring operation. In these cases, the name of the \emph{destination}
is usually a more descriptive name for the \emph{source}, while retaining
the \emph{source} token. Therefore, we do not flag these casts. We
also check if the \emph{source}'s type is a substring of the \emph{destination}'s
type. If the \emph{destination} type contains the \emph{source} type,
we assume that the two types have similar semantics and refrain from
flagging these casts. For instance, our tool would not flag the third
example from \prettyref{tbl:entropy-examples} because the \emph{source}
\lstinline*baz* can be found in the destination \lstinline+fooBaz+. 

We generate a ranked list for the named casts based on their conditional
entropy value in order to select identifiers where the expressions
in \emph{source} and \emph{destination} are disparate. This was done
to generate the corpus which is comprised of all four categories of
named casts: \lstinline*const_cast*, \lstinline*dynamic_cast*, \lstinline*reinterpret_cast*
and \lstinline*static_cast*.

\section{Evaluation\label{subsec:Evaluation}}

\begin{table*}[tp]
{\scriptsize{}}%
\begin{tabular}{>{\raggedright}p{1.9cm}>{\raggedright}p{2.9cm}>{\raggedleft}p{0.8cm}>{\raggedleft}p{0.6cm}>{\raggedleft}p{0.6cm}>{\raggedleft}p{0.3cm}>{\raggedleft}p{0.5cm}>{\raggedleft}p{0.01cm}>{\raggedleft}p{0.7cm}>{\raggedleft}p{0.6cm}>{\raggedleft}p{0.3cm}>{\raggedleft}p{0.4cm}>{\raggedleft}p{0.6cm}>{\raggedleft}p{0.6cm}}
\toprule 
{\scriptsize{}Name} &
{\scriptsize{}Description} &
{\scriptsize{}KLoc} &
\multicolumn{4}{c}{{\scriptsize{}Assignment expressions}} &
 &
\multicolumn{4}{c}{{\scriptsize{} Call expressions}} &
{\scriptsize{}Total} &
{\scriptsize{}Casts per }\tabularnewline
 &
 &
 &
{\scriptsize{}S} &
{\scriptsize{}R} &
{\scriptsize{}D} &
{\scriptsize{}C} &
 &
{\scriptsize{}S} &
{\scriptsize{}R} &
{\scriptsize{}D} &
{\scriptsize{}C} &
 &
{\scriptsize{}KLoc}\tabularnewline
\midrule 
{\scriptsize{}\href{https://chromium.googlesource.com/v8/v8.git/+/8200c5d117}{V8}} &
{\scriptsize{}JavaScript Engine} &
{\scriptsize{}1,359} &
{\scriptsize{}1,262} &
{\scriptsize{}1,649} &
{\scriptsize{}0} &
{\scriptsize{}8} &
 &
{\scriptsize{}1,592} &
{\scriptsize{}353} &
{\scriptsize{}0} &
{\scriptsize{}4} &
{\scriptsize{}4,868} &
{\scriptsize{}3.6}\tabularnewline
\addlinespace[0.15cm]
{\scriptsize{}\href{https://chromium.googlesource.com/chromium/src/+/689912289c/net/}{Net}} &
{\scriptsize{}Networking Protocols} &
{\scriptsize{}766} &
{\scriptsize{}616} &
{\scriptsize{}1,153} &
{\scriptsize{}0} &
{\scriptsize{}26} &
 &
{\scriptsize{}693} &
{\scriptsize{}770} &
{\scriptsize{}0} &
{\scriptsize{}15} &
{\scriptsize{}3,273} &
{\scriptsize{}4.3}\tabularnewline
\addlinespace[0.15cm]
{\scriptsize{}\href{https://chromium.googlesource.com/chromium/src/+/689912289c/gpu/}{gpu}} &
{\scriptsize{}Graphics Stack} &
{\scriptsize{}277} &
{\scriptsize{}1,386} &
{\scriptsize{}307} &
{\scriptsize{}0} &
{\scriptsize{}10} &
 &
{\scriptsize{}171} &
{\scriptsize{}100} &
{\scriptsize{}0} &
{\scriptsize{}56} &
{\scriptsize{}2,030} &
{\scriptsize{}7.3}\tabularnewline
\addlinespace[0.15cm]
{\scriptsize{}\href{https://chromium.googlesource.com/chromium/src/+/689912289c/ui/}{UI}} &
{\scriptsize{}UI Frameworks} &
{\scriptsize{}179} &
{\scriptsize{}197} &
{\scriptsize{}823} &
{\scriptsize{}0} &
{\scriptsize{}5} &
 &
{\scriptsize{}689} &
{\scriptsize{}36} &
{\scriptsize{}0} &
{\scriptsize{}4} &
{\scriptsize{}1,754} &
{\scriptsize{}9.8}\tabularnewline
\addlinespace[0.15cm]
{\scriptsize{}\href{https://chromium.googlesource.com/chromium/src/+/689912289c/media/}{Media}} &
{\scriptsize{}Media Components} &
{\scriptsize{}370} &
{\scriptsize{}450} &
{\scriptsize{}700} &
{\scriptsize{}0} &
{\scriptsize{}20} &
 &
{\scriptsize{}358} &
{\scriptsize{}207} &
{\scriptsize{}0} &
{\scriptsize{}3} &
{\scriptsize{}1,738} &
{\scriptsize{}4.7}\tabularnewline
\addlinespace[0.15cm]
{\scriptsize{}\href{https://chromium.googlesource.com/chromium/src/+/689912289c/third_party/blink/}{Blink}} &
{\scriptsize{}Browser Engine} &
{\scriptsize{}1,524} &
{\scriptsize{}1,081} &
{\scriptsize{}120} &
{\scriptsize{}0} &
{\scriptsize{}0} &
 &
{\scriptsize{}138} &
{\scriptsize{}0} &
{\scriptsize{}0} &
{\scriptsize{}0} &
{\scriptsize{}1,339} &
{\scriptsize{}0.9}\tabularnewline
\addlinespace[0.15cm]
{\scriptsize{}\href{https://chromium.googlesource.com/chromium/src/+/689912289c/chrome/}{Chrome}} &
{\scriptsize{}Application Layer} &
{\scriptsize{}2,385} &
{\scriptsize{}776} &
{\scriptsize{}199} &
{\scriptsize{}0} &
{\scriptsize{}22} &
 &
{\scriptsize{}256} &
{\scriptsize{}3} &
{\scriptsize{}0} &
{\scriptsize{}0} &
{\scriptsize{}1,256} &
{\scriptsize{}0.5}\tabularnewline
\addlinespace[0.15cm]
{\scriptsize{}\href{https://webrtc.googlesource.com/src.git/+/f1e97b9}{Webrtc}} &
{\scriptsize{}Communications API} &
{\scriptsize{}634} &
{\scriptsize{}482} &
{\scriptsize{}78} &
{\scriptsize{}0} &
{\scriptsize{}9} &
 &
{\scriptsize{}541} &
{\scriptsize{}33} &
{\scriptsize{}0} &
{\scriptsize{}1} &
{\scriptsize{}1,144} &
{\scriptsize{}1.8}\tabularnewline
\addlinespace[0.15cm]
{\scriptsize{}\href{https://skia.googlesource.com/skia.git/+/a1ea0a96f4}{Skia}} &
{\scriptsize{}Graphics Library} &
{\scriptsize{}665} &
{\scriptsize{}349} &
{\scriptsize{}274} &
{\scriptsize{}0} &
{\scriptsize{}20} &
 &
{\scriptsize{}208} &
{\scriptsize{}179} &
{\scriptsize{}0} &
{\scriptsize{}33} &
{\scriptsize{}1,063} &
{\scriptsize{}1.6}\tabularnewline
\addlinespace[0.15cm]
{\scriptsize{}\href{https://chromium.googlesource.com/chromium/src/+/689912289c/device/}{Device}} &
{\scriptsize{}Sensor Communication} &
{\scriptsize{}134} &
{\scriptsize{}469} &
{\scriptsize{}376} &
{\scriptsize{}0} &
{\scriptsize{}0} &
 &
{\scriptsize{}116} &
{\scriptsize{}30} &
{\scriptsize{}0} &
{\scriptsize{}0} &
{\scriptsize{}991} &
{\scriptsize{}7.4}\tabularnewline
\addlinespace[0.15cm]
{\scriptsize{}\href{https://chromium.googlesource.com/chromium/src/+/689912289c/chrome/browser/policy/}{Policy}} &
{\scriptsize{}Policy Settings} &
{\scriptsize{}39} &
{\scriptsize{}121} &
{\scriptsize{}34} &
{\scriptsize{}0} &
{\scriptsize{}353} &
 &
{\scriptsize{}314} &
{\scriptsize{}34} &
{\scriptsize{}0} &
{\scriptsize{}0} &
{\scriptsize{}856} &
{\scriptsize{}22.2}\tabularnewline
\addlinespace[0.15cm]
{\scriptsize{}\href{https://android.googlesource.com/platform/external/perfetto.git/+/b06d185a49}{Perfetto}} &
{\scriptsize{}Tracing Service} &
{\scriptsize{}205} &
{\scriptsize{}297} &
{\scriptsize{}7} &
{\scriptsize{}0} &
{\scriptsize{}54} &
 &
{\scriptsize{}454} &
{\scriptsize{}1} &
{\scriptsize{}0} &
{\scriptsize{}0} &
{\scriptsize{}813} &
{\scriptsize{}4}\tabularnewline
\addlinespace[0.15cm]
{\scriptsize{}\href{https://chromium.googlesource.com/chromium/src/+/689912289c/chrome/browser/safe_browsing/}{Safe Browse}} &
{\scriptsize{}URL Check Protocol} &
{\scriptsize{}9} &
{\scriptsize{}162} &
{\scriptsize{}57} &
{\scriptsize{}0} &
{\scriptsize{}79} &
 &
{\scriptsize{}440} &
{\scriptsize{}46} &
{\scriptsize{}0} &
{\scriptsize{}0} &
{\scriptsize{}784} &
{\scriptsize{}86.7}\tabularnewline
\addlinespace[0.15cm]
{\scriptsize{}\href{https://dawn.googlesource.com/dawn.git/+/0da52f2}{Dawn}} &
{\scriptsize{}WebGPU} &
{\scriptsize{}66} &
{\scriptsize{}125} &
{\scriptsize{}542} &
{\scriptsize{}0} &
{\scriptsize{}0} &
 &
{\scriptsize{}25} &
{\scriptsize{}3} &
{\scriptsize{}0} &
{\scriptsize{}0} &
{\scriptsize{}695} &
{\scriptsize{}10.5}\tabularnewline
\addlinespace[0.15cm]
{\scriptsize{}\href{https://chromium.googlesource.com/chromium/src/+/689912289c/third_party/protobuf/}{Protobuf}} &
{\scriptsize{}Serializing Struct Data} &
{\scriptsize{}227} &
{\scriptsize{}160} &
{\scriptsize{}77} &
{\scriptsize{}0} &
{\scriptsize{}17} &
 &
{\scriptsize{}394} &
{\scriptsize{}10} &
{\scriptsize{}0} &
{\scriptsize{}15} &
{\scriptsize{}673} &
{\scriptsize{}3}\tabularnewline
\addlinespace[0.15cm]
{\scriptsize{}\href{https://chromium.googlesource.com/chromium/src/+/689912289c/chrome/common/}{Common}} &
{\scriptsize{}Application Layer} &
{\scriptsize{}40} &
{\scriptsize{}341} &
{\scriptsize{}319} &
{\scriptsize{}0} &
{\scriptsize{}1} &
 &
{\scriptsize{}9} &
{\scriptsize{}0} &
{\scriptsize{}0} &
{\scriptsize{}0} &
{\scriptsize{}670} &
{\scriptsize{}16.8}\tabularnewline
\addlinespace[0.15cm]
{\scriptsize{}\href{https://chromium.googlesource.com/chromium/src/+/689912289c/base/}{Base}} &
{\scriptsize{}Core Components} &
{\scriptsize{}278} &
{\scriptsize{}192} &
{\scriptsize{}220} &
{\scriptsize{}0} &
{\scriptsize{}7} &
 &
{\scriptsize{}129} &
{\scriptsize{}102} &
{\scriptsize{}0} &
{\scriptsize{}6} &
{\scriptsize{}656} &
{\scriptsize{}2.4}\tabularnewline
\addlinespace[0.15cm]
{\scriptsize{}\href{https://pdfium.googlesource.com/pdfium/+/0f4ac587a4}{Pdfium}} &
{\scriptsize{}PDF Library} &
{\scriptsize{}484} &
{\scriptsize{}369} &
{\scriptsize{}62} &
{\scriptsize{}0} &
{\scriptsize{}1} &
 &
{\scriptsize{}181} &
{\scriptsize{}20} &
{\scriptsize{}0} &
{\scriptsize{}0} &
{\scriptsize{}633} &
{\scriptsize{}1.3}\tabularnewline
\addlinespace[0.15cm]
{\scriptsize{}\href{https://chromium.googlesource.com/chromium/deps/icu/+/2ecd66c696/}{ICU}} &
{\scriptsize{}Unicode Components} &
{\scriptsize{}325} &
{\scriptsize{}285} &
{\scriptsize{}63} &
{\scriptsize{}75} &
{\scriptsize{}40} &
 &
{\scriptsize{}79} &
{\scriptsize{}14} &
{\scriptsize{}1} &
{\scriptsize{}5} &
{\scriptsize{}562} &
{\scriptsize{}1.7}\tabularnewline
\addlinespace[0.15cm]
{\scriptsize{}\href{https://chromium.googlesource.com/chromium/src/+/689912289c/components/viz/}{VIZ}} &
{\scriptsize{}Visual Subservices} &
{\scriptsize{}84} &
{\scriptsize{}176} &
{\scriptsize{}235} &
{\scriptsize{}0} &
{\scriptsize{}0} &
 &
{\scriptsize{}51} &
{\scriptsize{}57} &
{\scriptsize{}0} &
{\scriptsize{}0} &
{\scriptsize{}519} &
{\scriptsize{}6.2}\tabularnewline
\addlinespace[0.15cm]
{\scriptsize{}\href{https://chromium.googlesource.com/chromium/src/+/689912289c/components/metrics}{Metrics Proto}} &
{\scriptsize{}Data Analysis} &
{\scriptsize{}75} &
{\scriptsize{}165} &
{\scriptsize{}0} &
{\scriptsize{}0} &
{\scriptsize{}47} &
 &
{\scriptsize{}304} &
{\scriptsize{}0} &
{\scriptsize{}0} &
{\scriptsize{}0} &
{\scriptsize{}516} &
{\scriptsize{}6.9}\tabularnewline
\addlinespace[0.15cm]
{\scriptsize{}\href{https://chromium.googlesource.com/chromium/src/+/689912289c/chrome/browser/sync/}{Sync}} &
{\scriptsize{}Sync Implementation} &
{\scriptsize{}140} &
{\scriptsize{}92} &
{\scriptsize{}1} &
{\scriptsize{}0} &
{\scriptsize{}84} &
 &
{\scriptsize{}313} &
{\scriptsize{}3} &
{\scriptsize{}0} &
{\scriptsize{}0} &
{\scriptsize{}493} &
{\scriptsize{}3.5}\tabularnewline
\addlinespace[0.15cm]
{\scriptsize{}\href{https://chromium.googlesource.com/angle/angle.git/+/2328d65ab}{Angle}} &
{\scriptsize{}Graphics Engine} &
{\scriptsize{}2,381} &
{\scriptsize{}175} &
{\scriptsize{}28} &
{\scriptsize{}0} &
{\scriptsize{}3} &
 &
{\scriptsize{}230} &
{\scriptsize{}19} &
{\scriptsize{}0} &
{\scriptsize{}0} &
{\scriptsize{}455} &
{\scriptsize{}0.2}\tabularnewline
\addlinespace[0.15cm]
{\scriptsize{}\href{https://chromium.googlesource.com/chromium/src/+/689912289c/buildtools/}{Buildtools}} &
{\scriptsize{}Buildtools Chromium} &
{\scriptsize{}510} &
{\scriptsize{}187} &
{\scriptsize{}153} &
{\scriptsize{}13} &
{\scriptsize{}2} &
 &
{\scriptsize{}25} &
{\scriptsize{}7} &
{\scriptsize{}0} &
{\scriptsize{}3} &
{\scriptsize{}390} &
{\scriptsize{}0.8}\tabularnewline
\addlinespace[0.15cm]
{\scriptsize{}\href{https://chromium.googlesource.com/chromium/src/+/689912289c/media/audio/}{Audio}} &
{\scriptsize{}Audio System} &
{\scriptsize{}34} &
{\scriptsize{}43} &
{\scriptsize{}202} &
{\scriptsize{}0} &
{\scriptsize{}0} &
 &
{\scriptsize{}33} &
{\scriptsize{}50} &
{\scriptsize{}0} &
{\scriptsize{}0} &
{\scriptsize{}328} &
{\scriptsize{}9.6}\tabularnewline
\addlinespace[0.15cm]
{\scriptsize{}\href{https://swiftshader.googlesource.com/SwiftShader.git/+/0cd9a67ce}{Swiftshader}} &
{\scriptsize{}Graphics Library} &
{\scriptsize{}2,166} &
{\scriptsize{}160} &
{\scriptsize{}87} &
{\scriptsize{}0} &
{\scriptsize{}5} &
 &
{\scriptsize{}62} &
{\scriptsize{}6} &
{\scriptsize{}0} &
{\scriptsize{}0} &
{\scriptsize{}320} &
{\scriptsize{}0.1}\tabularnewline
\addlinespace[0.15cm]
{\scriptsize{}\href{https://chromium.googlesource.com/chromium/src/+/689912289c/extensions/}{Extensions}} &
{\scriptsize{}Core Parts Extension} &
{\scriptsize{}224} &
{\scriptsize{}312} &
{\scriptsize{}4} &
{\scriptsize{}0} &
{\scriptsize{}0} &
 &
{\scriptsize{}0} &
{\scriptsize{}0} &
{\scriptsize{}0} &
{\scriptsize{}0} &
{\scriptsize{}316} &
{\scriptsize{}1.4}\tabularnewline
\addlinespace[0.15cm]
{\scriptsize{}\href{https://chromium.googlesource.com/chromium/src/+/689912289c/cc/}{CC}} &
{\scriptsize{}Compositor Renderer} &
{\scriptsize{}198} &
{\scriptsize{}117} &
{\scriptsize{}17} &
{\scriptsize{}0} &
{\scriptsize{}0} &
 &
{\scriptsize{}167} &
{\scriptsize{}6} &
{\scriptsize{}0} &
{\scriptsize{}2} &
{\scriptsize{}309} &
{\scriptsize{}1.6}\tabularnewline
\addlinespace[0.15cm]
{\scriptsize{}\href{https://chromium.googlesource.com/chromium/src/+/689912289c/components/remote_cocoa/}{Remote Cocoa}} &
{\scriptsize{}Cocoa Front-End} &
{\scriptsize{}4} &
{\scriptsize{}137} &
{\scriptsize{}158} &
{\scriptsize{}0} &
{\scriptsize{}0} &
 &
{\scriptsize{}5} &
{\scriptsize{}1} &
{\scriptsize{}0} &
{\scriptsize{}0} &
{\scriptsize{}301} &
{\scriptsize{}70.7}\tabularnewline
\addlinespace[0.15cm]
{\scriptsize{}\href{https://chromium.googlesource.com/chromium/src/+/689912289c/base/}{Logging}} &
{\scriptsize{}Logs Implementation} &
{\scriptsize{}43} &
{\scriptsize{}90} &
{\scriptsize{}0} &
{\scriptsize{}0} &
{\scriptsize{}6} &
 &
{\scriptsize{}176} &
{\scriptsize{}0} &
{\scriptsize{}0} &
{\scriptsize{}0} &
{\scriptsize{}272} &
{\scriptsize{}6.3}\tabularnewline
\addlinespace[0.15cm]
{\scriptsize{}Rest of Corpus} &
{\scriptsize{}Components < 250} &
{\scriptsize{}11,175} &
{\scriptsize{}2,238} &
{\scriptsize{}1,284} &
{\scriptsize{}0} &
{\scriptsize{}247} &
 &
{\scriptsize{}1,925} &
{\scriptsize{}545} &
{\scriptsize{}0} &
{\scriptsize{}42} &
{\scriptsize{}6,281} &
{\scriptsize{}0.6}\tabularnewline
\midrule 
 &
\textbf{\scriptsize{}Total Casts} &
{\scriptsize{}27,000} &
{\scriptsize{}13,012} &
{\scriptsize{}9,229} &
{\scriptsize{}88} &
{\scriptsize{}1,066} &
 &
{\scriptsize{}10,078 } &
{\scriptsize{}2,635} &
{\scriptsize{}1} &
{\scriptsize{}189} &
{\scriptsize{}36,298} &
{\scriptsize{}1.3}\tabularnewline
\bottomrule
\addlinespace[0.2cm]
\end{tabular}{\scriptsize\par}

\caption{C++ Corpus from Google Chromium. Represents the distribution of cast
types and the frequency of usage of each conversion operator (S -
\lstinline*static_cast*, R - \lstinline*reinterpret_cast*, D - \lstinline*dynamic_cast*,
C - \lstinline*const_cast*)}

\label{tbl:dataset}\vspace{-15bp}
\end{table*}

We now present the results of our evaluation of our tool on the Chromium
corpus. Through our experiments, we try to address the following research
questions.

\paragraph{RQ\protect\textsubscript{1}}

How prevalent is the usage of named casts in large and mainstream
open-source projects such as the Chromium project? (\hyperref[subsec:Quantitative-analysis]{Section 4.1})

\paragraph*{RQ\protect\textsubscript{2}}

Is our tool effective at identifying poor practices, in both how casts
are used and how identifiers in the casts are named? (\hyperref[subsec:pr-curve]{Section 4.4})

\paragraph*{RQ\protect\textsubscript{3}}

Are longer \emph{source} expressions, which are harder to comprehend,
correlated with poor cast usage? (\hyperref[subsec:pr-curve]{Section 4.5})

\paragraph*{RQ\protect\textsubscript{4}}

What is the nature of some of the poor practices in choosing names
and usage of casts identified by the tool? (\hyperref[subsec:Qualitative-analysis]{Section 4.6})

The goal for \textbf{RQ}\textsubscript{\textbf{1}} is to assess
how often the named cast operators are used in large projects. This
helps to identify if there would be a need for a tool to validate
named cast operators. \textbf{RQ}\textsubscript{\textbf{2}} checks
that the collected named cast operations can be validated using the
identifiers information. For \textbf{RQ}\textsubscript{\textbf{3}},
we look into long \emph{source} expressions (those of 100+ characters)
which are harder to comprehend by developers.\textbf{ RQ}\textsubscript{\textbf{3}}
tries to observe if the relation between \emph{source} length and
conditional entropy value can lead to detecting casts misuse.\textbf{
}The scope of \textbf{RQ}\textsubscript{\textbf{4}} is to generate
insights about each named cast operator's usages and the identifier
names used.

\subsection{Quantitative Analysis (RQ\protect\textsubscript{\textbf{1}}) \label{subsec:Quantitative-analysis}}

To answer \textbf{RQ}\textsubscript{\textbf{1}}, we present the
usage patterns of named cast operators in Chromium's components. \prettyref{tbl:dataset}
shows the distribution of the named casts in various components of
Chromium. Our corpus consists of 36,298 named casts. \prettyref{tbl:dataset}
shows the frequency for each category of named casts for individual
modules in the Chromium corpus. Overall, 63.62\% are \lstinline*static_cast*s,
32.68\% are \lstinline*reinterpret_cast*s, 0.25\% are \lstinline*dynamic_cast*s
and 3.45\% are \lstinline*const_cast*s. As discussed in \hyperref[subsec:Extraction-of-Named-Casts]{Section 3.2},
we consider named casts that are a part of either assignments or actual-to-formal
parameter binding in function calls. The proportion of named casts
that are a part of assignments is 64.46\% (23,395 casts) while 35.54\%
(12,903 casts) are in call expressions. Overall, there are 1.3 named
cast operations per one thousand lines of code across Chromium. \prettyref{tbl:dataset}
shows that the number of casts per KLOC varies and it can take values
from 0.1 up to 86.7 depending on the module's scope. This shows that
some developers, particularly those on the higher end of cast usage
that we observed, could benefit from a tool to vet their named cast
operations.

It is observed from \prettyref{tbl:dataset} that \lstinline*dynamic_cast*
and \lstinline*const_cast* operators are used rarely. The \lstinline*dynamic_cast*
operator uses Run-Time Type Identification (RTTI) to verify that the
types can be converted at runtime, which is an expensive operation.
It is likely that the cost of checking prohibits their widespread
use. \lstinline*const_cast* operators are used to set or remove the
constness or volatility of variables. Such variables are rare themselves
which explains why so few instances of \lstinline*const_cast* are
present in our dataset. \lstinline*static_cast* can be used to cast
up or down objects. A check on the class inheritance hierarchy evaluates
if the conversion between the object and \emph{destination} type is
possible. Therefore, \lstinline*static_cast* is safer than \lstinline*reinterpret_cast*
which is extremely permissive, allowing arbitrary type conversions.
Indeed, best practice is to use \lstinline*static_cast* over \lstinline*reinterpret_cast*
and this is reflected in the prevalence of \lstinline*static_cast*
operations in our corpus. It is noticed from \prettyref{tbl:dataset}
that the larger and performance-critical modules such as the JavaScript
compiler \emph{V8}, networking (\emph{Net}), GPU, user interface (\emph{UI}),
the \emph{Media} libraries, etc. have the most casts. Interestingly,
none of these modules uses the runtime intensive \lstinline*dynamic_cast*
cast operators. Only International Components for Unicode\emph{ (ICU)}
and \emph{Buildtools} components contain a total of 88 \lstinline*dynamic_cast*
operators. Neither of these components are central to the user experience
of the browser and thus they can potentially tolerate runtime overheads.

\subsection{Establishing Ground Truth \label{subsec:manual-evaluation}}

We performed a manual investigation of a random-uniform sample of
the corpus to establish the ground truth. The ground truth is essential
to present trade-offs at different thresholds above which casts are
flagged by our tool. Using the central limit theorem {[}\citeauthor{central_limit_theorem}
\citeyear{central_limit_theorem}{]}, we targeted a 90\% confidence
with a 5\% margin of error for the sampled set. The set comprised
of 271 data points with a breakdown of 163 \lstinline*static_cast*,
97 \lstinline*reinterpret_cast*, 11 \lstinline*const_cast* and 0
\lstinline*dynamic_cast* operations. There are no \lstinline*dynamic_cast*
operations in the sampled set because the casts corpus contains only
89 of these cases out of 36,298, thus the probability of randomly
selecting any of these cases is small. 

\paragraph*{Raters}

We used three raters to establish the ground truth. All raters were
male residents of the United Kingdom. Their occupations were: full-time
undergraduate student (R1), full-time PhD student (R2) and Lecturer
(R3), with 3, 7 and 20 years of programming experience respectively.
The raters were provided with full access to the \emph{source} code
for the casts to mark casts as true or false. The true cases represent
casts that are poorly implemented or they have imprecise names for
identifiers. In contrast, false cases are correct and efficient implementations
with meaningful names for the identifiers. 

\paragraph*{Guidelines and Protocol}

The raters were looking for instances when the casts were misused
or part of inefficient or buggy code. To understand if identifiers
are descriptive and concordant with their meaning, the raters were
advised to look at usages of the variable as well as statements before
and after the cast expression. To identify poor implementation of
casts, the raters inspected the code that was referring to the cast,
the variables\textquoteright{} definitions, the variables\textquoteright{}
usages, the user-defined type definitions and the git logs for changes
related to the casts. The raters consulted ISO C++ standards {[}\citeauthor{iso_cpp}
\citeyear{iso_cpp}{]} to revise their understandings of each named
cast operator usage. 

\paragraph*{Inter-Rater Agreement}

Overall, the human evaluators classified on average 71 out of 271
cases as being true noting that 26.9\% of cases as casts were misuses
and 73.1\% as cases with imprecise names. The individual ratings are
73, 71 and 71 for R1, R2 and R3, respectively. The inter-rater agreement,
also called Cohen's Kappa coefficient {[}\citeauthor{kappa} \citeyear{kappa}{]},
was used to measure agreement level. Kappa can take values between
-1 and 1. If Kappa has a value of 1, it means that the raters are
in perfect agreement. A negative Kappa value means that the raters
are in disagreement. Kappa coefficient is calculated as the mean value
between the kappa coefficient between any two raters. The Cohen's
Kappa coefficient for this evaluation between raters had the following
values: 0.85, 0.89 and 0.96. This means that raters had high levels
of agreement between them. 

\begin{table*}
{\footnotesize{}}%
\subfloat[Degrees of Freedom. For each degree of freedom, a value {*} is used
if any value can be used.]%
{{\footnotesize{}}%
\begin{tabular*}{1\textwidth}{@{\extracolsep{\fill}}>{\raggedright}p{0.1\textwidth}>{\raggedright}p{0.4\textwidth}>{\raggedright}p{0.4\textwidth}}
\hline 
\textbf{\footnotesize{}Attributes} &
\textbf{\footnotesize{}Description} &
\textbf{\footnotesize{}Values}\tabularnewline
\hline 
{\footnotesize{}Cast Type} &
{\footnotesize{}The type of named cast used} &
{\footnotesize{}}{\footnotesize{}\lstinline*static_cast*}{\footnotesize{},
}{\footnotesize{}\lstinline*dynamic_cast*}{\footnotesize{},}{\footnotesize{}\lstinline*const_cast*}{\footnotesize{},
}{\footnotesize{}\lstinline*reinterpret_cast*}\tabularnewline[0.5cm]
{\footnotesize{}Identifiers} &
{\footnotesize{}Usage of meaningful identifiers indicative of semantics} &
{\footnotesize{}\{Specific (S), Generic (G), Value (V) eg. literals,
Not Indicative/Imprecise (N)\}}\tabularnewline[0.5cm]
{\footnotesize{}Direction} &
{\footnotesize{}Whether a cast is a value or expression is being upcast
or downcast} &
{\footnotesize{}\{Downcast (D), Upcast(U), Coercion-like conversion
(C)\}}\tabularnewline[0.5cm]
\hline 
\end{tabular*}{\footnotesize\par}

{\footnotesize{}\label{tbl:taxonomy_freedom_degrees}}{\footnotesize\par}}{\footnotesize\par}

{\footnotesize{}}%
\subfloat[Categorisation of Casts. The values within the tuple < , > represents
the expressiveness of \emph{source} and \emph{destination}.]%
{{\footnotesize{}}%
\begin{tabular*}{1\textwidth}{@{\extracolsep{\fill}}>{\raggedright}m{2cm}l>{\raggedright}m{0.3\textwidth}>{\raggedright}m{0.2\textwidth}}
\hline 
\textbf{\footnotesize{}Category} &
\textbf{\footnotesize{}Combination} &
\textbf{\footnotesize{}Description} &
\textbf{\footnotesize{}Applications or Examples}\tabularnewline
\hline 
{\footnotesize{}Assign Value} &
{\footnotesize{}{*} , <V, S>, C} &
{\footnotesize{}Assign values, generally of primitive type, to variables,
pointers or objects} &
{\footnotesize{}Assign default values (min, max, size, etc.), calculation
results, bitwise operations, etc.}\tabularnewline[1.1cm]
{\footnotesize{}Conversions Between Classes} &
\multirow{1}{*}[0.45cm]{{\footnotesize{}{*}, <{*}, {*}>, \{D, U\}}} &
{\footnotesize{}Objects or pointer of objects being upcast, downcast
or the bytes being reinterpreted in related class types} &
{\footnotesize{}Access of related class methods or members.}\tabularnewline[1.1cm]
{\footnotesize{}Specialisation} &
{\footnotesize{}{*} , <{*}, {*}>, D} &
{\footnotesize{}A generic pointer, of the type }\lstinline+void*+{\footnotesize{},
being cast into a type with more specific properties} &
{\footnotesize{}Initiliasation of iterators.}\tabularnewline[1.1cm]
{\footnotesize{}Rapid Prototyping} &
\multirow{1}{*}[0.25cm]{{\footnotesize{}{*} , <{*}, {*}>, C}} &
\multirow{1}{0.3\textwidth}[0.25cm]{{\footnotesize{}These cases indicate the casts are being used to cross
type boundaries for Software Engineering reasons}} &
{\footnotesize{}Serialisation or deserialisation of objects, or e.g.
}{\footnotesize{}\lstinline*fileBuffer*}{\footnotesize{} to }{\footnotesize{}\lstinline*networkBuffer*}\tabularnewline[1.1cm]
{\footnotesize{}Modifiability} &
{\footnotesize{}}{\footnotesize{}\lstinline*const_cast*}{\footnotesize{}
, <{*}, {*}>, {*}} &
{\footnotesize{}Modifying the }{\footnotesize{}\lstinline*const*}{\footnotesize{}
or }{\footnotesize{}\lstinline*volatile*}{\footnotesize{} property
of a variable} &
{\footnotesize{}Calling non-}{\footnotesize{}\lstinline*const*}{\footnotesize{}
functions on }{\footnotesize{}\lstinline*const*}{\footnotesize{}
variables e.g. }{\footnotesize{}\lstinline*print*}\tabularnewline[1.1cm]
{\footnotesize{}Check and Cast} &
\multirow{1}{*}[0.25cm]{{\footnotesize{}}{\footnotesize{}\lstinline*dynamic_cast*}{\footnotesize{}
, <{*}, {*}>, {*}}} &
\multirow{1}{0.3\textwidth}[0.25cm]{{\footnotesize{}The }{\footnotesize{}\lstinline*dynamic_cast*}{\footnotesize{}
operator ensures that the source can be converted to the destination
type}} &
{\footnotesize{}Validate user input and rely on the operator for safety}\tabularnewline[1.1cm]
\multirow{2}{2cm}[-0.5cm]{{\footnotesize{}Imprecise Naming}} &
{\footnotesize{}{*} , \{<N, {*}>, <{*}, N>\}, {*}} &
{\footnotesize{}Any case where the source or destination does not
sufficiently describe the data or the operation, which can cause confusion} &
\multirow{2}{0.2\textwidth}[-0.5cm]{{\footnotesize{}Some examples are presented in }{\footnotesize{}\hyperref[subsec:Motivation]{Section 2.3}}}\tabularnewline[1cm]
 & {\footnotesize{}{*} , <G, G>, {*}} &
{\footnotesize{}Any case where the source or destination is too generic
and does not improve the clarity of the code} & \tabularnewline
\hline 
\end{tabular*}{\footnotesize\par}

{\footnotesize{}\label{tbl:taxonomy_usage}}{\footnotesize\par}}{\footnotesize\par}

{\small{}\label{tbl:taxonomy_categories}}\caption{A Taxonomy of Casts.}
\end{table*}

\subsection{Cast Taxonomy\label{subsec:cast_taxonomy}}

The usage of casts can vary across projects, but for every cast operation
there are a few degrees of freedom. These include, but are not limited
to, the type of the cast operator, the use of variables or expressions
in the \emph{source} and \emph{destination} and the names of these
variables, as well as the tokens in the expressions. We also consider
the direction of the cast as part of the degrees of freedom. The directions
of the cast can be Upcast or Downcast when there is a hierarchical
class relation between the variables. In addition, there will also
be Coercion-like conversions between variables with no class relation,
for example, cast between numerical values. The degrees of freedom
are presented in \hyperref[tbl:taxonomy_freedom_degrees]{Table}\prettyref{tbl:taxonomy_freedom_degrees}.
Our research hypothesis is that identifiers used for variable names
and those in expressions carry semantics. Therefore, in our research,
and subsequently, our taxonomy, we treat choice of identifiers as
a degree of freedom. In addition, whether a cast is being used to
upcast or downcast the \emph{source} potentially influences the cast's
role in the wider code; hence, in our categorisation, we treat this
as an input.

Named cast operations have many applications, and we have split these
into six major categories: Assign Values, Conversions Between Classes,
Specialisation, Rapid Prototyping, Modifiability and Check and Cast.
As it can be seen from the combination of inputs, Specialisation category
is a more specific usage of Conversion Between Classes of \lstinline+void*+
pointers. On top of the six categories, there is the additional negative
category, Imprecise Naming, which we capture through our tool. We
established these categories after consulting background information
about each named cast operator, as presented in \hyperref[subsec:Type-conversions]{Section 2.1}.
This is in addition to careful consideration following first-hand
observations of cast usage during our manual evaluation of the Chromium
corpus. \hyperref[tbl:taxonomy_usage]{Table}\prettyref{tbl:taxonomy_usage}
presents the combination of the degrees of freedom for each category,
along with a description and some applications. 

\paragraph*{Casts Usage in the Ground Truth Set}

To better understand the ground truth set, the raters grouped the
named cast operations using the taxonomy classification. \prettyref{tbl:taxonomy}
presents the frequency of named cast cases for each category. To provide
a better description of the ground truth set, some additional applications
have been added to the following categories: Assign Value and Conversion
Between Classes. Out of the 271 cases, 121 are in Assign Value, 53
are in Conversion Between Classes, 64 are in Specialisation, 23 are
in Prototyping, 10 are in Modifiability and none are in Check and
Cast. Of all the cases, only 71 cases belong in the Imprecise Names
category. Based on the identifiers' quality, there are 138 cases where
\emph{source} and \emph{destination} are specific, 20 cases where
the \emph{source} is a value and \emph{destination} is specific, 50
cases of generic identifiers and 64 cases where the identifiers are
not indicative. 

\begin{table}
{\footnotesize{}}%
\begin{tabular}{>{\raggedright}p{5cm}c}
\hline 
{\footnotesize{}Categories of named cast usage} &
{\footnotesize{}\# of cases}\tabularnewline
\hline 
\multicolumn{2}{l}{\textbf{\footnotesize{}Assign Values}}\tabularnewline
{\footnotesize{}Create a local copy of a value} &
{\footnotesize{}50}\tabularnewline
{\footnotesize{}Default values (max, min, size, etc.)} &
{\footnotesize{}46}\tabularnewline
{\footnotesize{}Value calculations result } &
{\footnotesize{}10}\tabularnewline
{\footnotesize{}Enumerations conversions} &
{\footnotesize{}9}\tabularnewline
{\footnotesize{}Bitwise operations result} &
{\footnotesize{}4}\tabularnewline
{\footnotesize{}Assign default address} &
{\footnotesize{}2}\tabularnewline
 &
\tabularnewline
\multicolumn{2}{l}{\textbf{\footnotesize{}Conversion between Classes}}\tabularnewline
{\footnotesize{}Address to address or pointer to pointer conversions} &
{\footnotesize{}34}\tabularnewline
{\footnotesize{}Upcast/Downcast} &
{\footnotesize{}12}\tabularnewline
{\footnotesize{}From address to specific data structure} &
{\footnotesize{}7}\tabularnewline
 &
\tabularnewline
\multicolumn{2}{l}{\textbf{\footnotesize{}Specialisation}}\tabularnewline
{\footnotesize{}Void type pointers conversions} &
{\footnotesize{}53}\tabularnewline
{\footnotesize{}Iterator Conversions} &
{\footnotesize{}11}\tabularnewline
 &
\tabularnewline
\multicolumn{2}{l}{\textbf{\footnotesize{}Rapid Prototyping}}\tabularnewline
{\footnotesize{}Serialisation/Deserialisation of objects} &
{\footnotesize{}23}\tabularnewline
 &
\tabularnewline
\multicolumn{2}{l}{\textbf{\footnotesize{}Modifiability}}\tabularnewline
{\footnotesize{}Modifying const property of a variable} &
{\footnotesize{}10}\tabularnewline
 &
\tabularnewline
\textbf{\footnotesize{}Check and Cast} &
{\footnotesize{}0}\tabularnewline
\textbf{\footnotesize{}Imprecise Naming} &
{\footnotesize{}71}\tabularnewline
\hline 
 &
\tabularnewline
 &
\tabularnewline
\hline 
{\footnotesize{}Categories of identifiers' quality} &
{\footnotesize{}\# of cases}\tabularnewline
\hline 
\multicolumn{2}{l}{\textbf{\footnotesize{}Specific/Precise Identifiers}}\tabularnewline
\emph{\footnotesize{}Source}{\footnotesize{} and }\emph{\footnotesize{}destination}{\footnotesize{}
are related} &
{\footnotesize{}125}\tabularnewline
\emph{\footnotesize{}Source}{\footnotesize{} and }\emph{\footnotesize{}destination}{\footnotesize{}
are the same} &
{\footnotesize{}13}\tabularnewline
 &
\tabularnewline
\multicolumn{2}{l}{\textbf{\emph{\footnotesize{}Source}}\textbf{\footnotesize{} is value
and }\textbf{\emph{\footnotesize{}destination}}\textbf{\footnotesize{}
is specific}}\tabularnewline
\emph{\footnotesize{}Source}{\footnotesize{} is an operation } &
{\footnotesize{}11}\tabularnewline
\emph{\footnotesize{}Source}{\footnotesize{} is a value of a primitive
type} &
{\footnotesize{}9}\tabularnewline
 &
\tabularnewline
\multicolumn{2}{l}{\textbf{\footnotesize{}Generic Identifiers}}\tabularnewline
{\footnotesize{}Identifiers are generic} &
{\footnotesize{}50}\tabularnewline
 &
\tabularnewline
\multicolumn{2}{l}{\textbf{\footnotesize{}Not indicative/imprecise Identifiers}}\tabularnewline
\emph{\footnotesize{}Source}{\footnotesize{} and }\emph{\footnotesize{}destination}{\footnotesize{}
are unrelated } &
{\footnotesize{}13}\tabularnewline
{\footnotesize{}Identifiers are generic} &
{\footnotesize{}51}\tabularnewline
\hline 
\end{tabular}{\footnotesize{} }{\footnotesize\par}

\caption{Sampled dataset quantified based on\protect \\
the taxonomy}
\label{tbl:taxonomy}
\end{table}

\subsection{Precision-Recall Curve (RQ\protect\textsubscript{2}) \label{subsec:pr-curve}}

\begin{figure*}[t]
\subfloat[Precision, Recall, Accuracy and F1 scores \protect \\
at different thresholds using both preprocessing filters.]%
{\includegraphics[width=1\columnwidth]{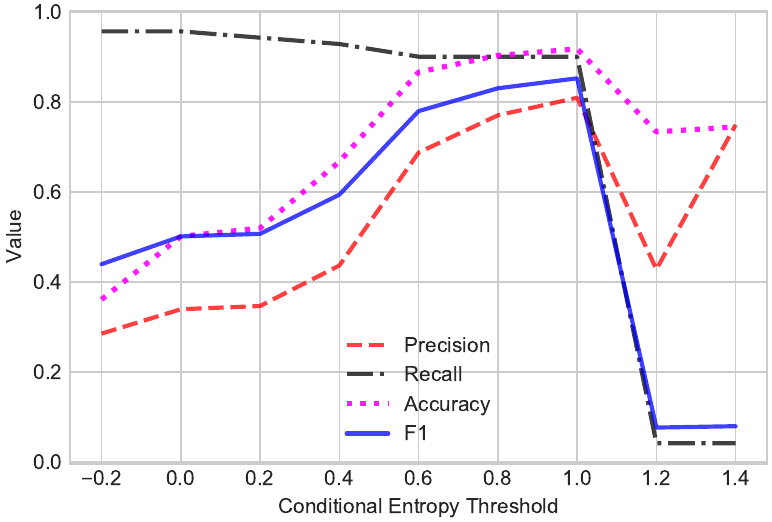}
\begin{centering}
\par\end{centering}
\label{fig:grouped_metrics}}%
\subfloat[Comparison of accuracy scores at different thresholds of the tool
with different settings for filters.]%
{\includegraphics[width=1\columnwidth]{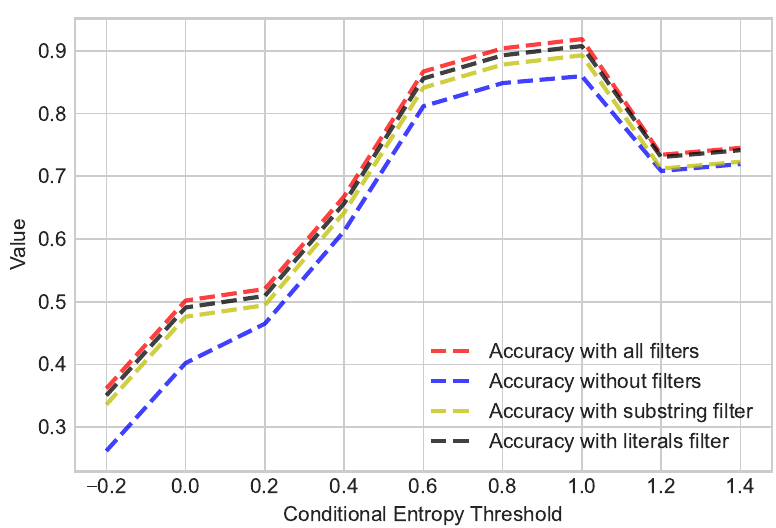}

\label{fig:grouped_accuracy}}

\subfloat[Comparison of precision scores at different thresholds\protect \\
 of the tool with different settings for filters.]%
{\includegraphics[width=1\columnwidth]{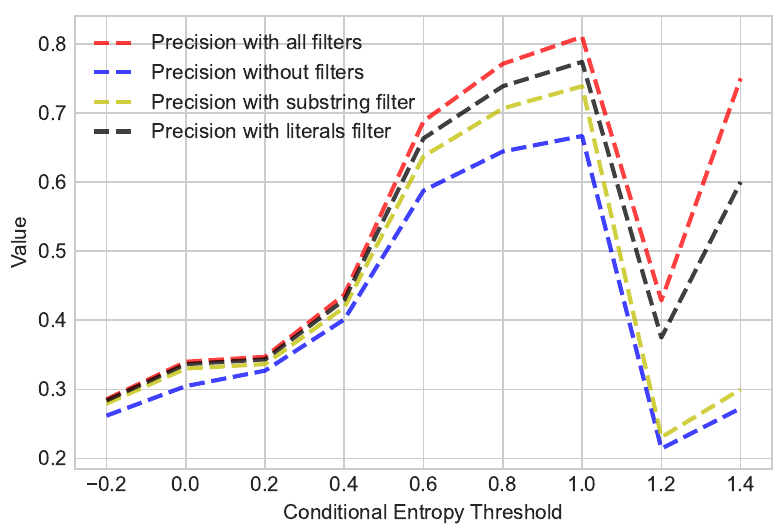}

\label{fig:grouped_precision}}%
\subfloat[Comparison of recall scores at different thresholds of \protect \\
the tool with different settings for filters.]%
{\includegraphics[width=1\columnwidth]{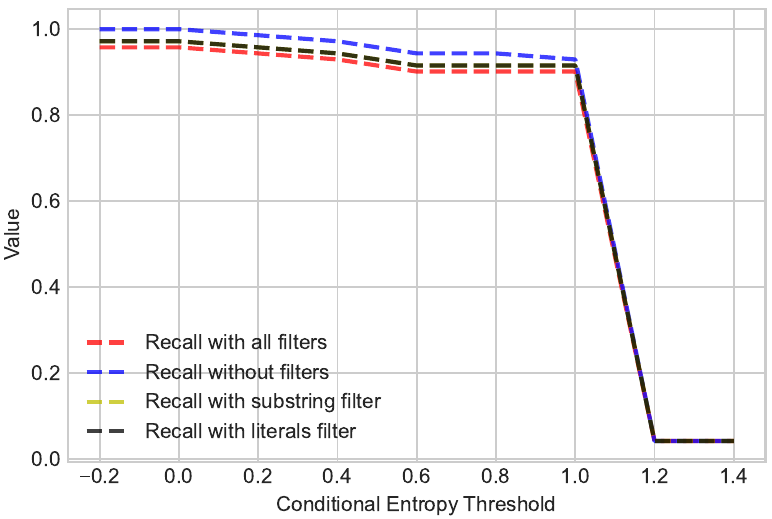}
\begin{centering}
\par\end{centering}
\label{fig:grouped_recall}}\caption{Tool's performance under different settings.}
\end{figure*}

To answer \textbf{RQ}\textsubscript{\textbf{2}}\textbf{,} we established
a ground truth on a sample dataset in order to quantify the results
of our tool. We investigate the optimal setting for the conditional
entropy threshold and report the tool's performance. For the sampled
dataset, the conditional entropy ranges from -0.08 to 2.58, with two
cases where the conditional entropy was 1.58 and above. The Conditional
Entropy is negative for cases where the joint entropy is smaller than
the \emph{source}'s entropy because the \emph{source} subtokens can
be found in the \emph{destination}, while the \emph{destination} is
unique. To optimise the threshold we selected values between -0.2
and 1.4 in increments of 0.2. For each level, we computed the precision,
recall, accuracy and F1-score. 

We first give the definition of our metrics before presenting our
results. Precision, presented in \prettyref{eq:precision}, is the
proportion of the actual flagged cases from all the cases marked as
flagged. \prettyref{eq:recall} shows the recall, which is the proportion
of flagged cases that our tool detects from all actual flagged cases.
Accuracy is presented in \prettyref{eq:accuracy} and it represents
the percentage of cases that our tool correctly classifies. F-measure
is the harmonic mean between precision and recall {[}\citeauthor{f-measure}
\citeyear{f-measure}{]}. For our study, we utilise the F1-score presented
in \prettyref{eq:f1} to evaluate the tool's performance.

\begin{equation}
Precision=\frac{TP}{TP+FP}\label{eq:precision}
\end{equation}
\begin{equation}
Recall=\frac{TP}{TP+FN}\label{eq:recall}
\end{equation}

\begin{equation}
Accuracy=\frac{TP+TN}{TP+TN+FP+FN}\label{eq:accuracy}
\end{equation}

\begin{equation}
F_{1}score=2\times\frac{Precision\times Recall}{Precision+Recall}\label{eq:f1}
\end{equation}

\hyperref[fig:grouped_metrics]{Figure}\prettyref{fig:grouped_metrics}
shows the precision, recall, accuracy and F1-score based on the different
threshold levels. The conditional entropy threshold represents the
value when our tool starts marking a case as true. At the lowest threshold
value, our tool marks all cases as true except those marked false
by the preprocessing filters. With the increase of the threshold value,
the precision and the F1-score also increase, while the recall decreases.
When the threshold takes the value of 1.2, the F1-score and recall
begin to fall, and the precision halves. This phenomenon happens because
the casts are distributed based on the conditional entropy and the
high number of true cases between the threshold values of 1 and 1.2.
In the corpus, there are only 1,405 cases with a conditional entropy
higher than 1.2. Thus, when the sampled dataset was collected, only
14 cases were selected with a conditional entropy higher than 1.2.
Based on the F1-score, we select the \textbf{value 1} as the \textbf{conditional
entropy threshold} for our tool. \hyperref[fig:grouped_metrics]{Figure}\prettyref{fig:grouped_metrics}
shows that, on the sampled dataset of 271 cases, the tool has a peak
precision of 81\%, a recall of 90\% and an accuracy of 92\% in identifying
correctly flagged cases. 

\paragraph{Impact of Filters}

\hyperref[fig:grouped_metrics]{Figures} \ref{fig:grouped_accuracy},
\ref{fig:grouped_precision} and \ref{fig:grouped_recall} show the
difference between precision, recall and accuracy for the tool with
and without the preprocessing filters. The use of preprocessing filters
improves the Precision value from 0.67 up to 0.81, and Accuracy from
0.86 up to 0.92. However, the recall is decreased from 0.93 to 0.9.
The Cast from Literals filter selects 9 cases and the Descriptive
Variables and Type filter selects another 25 cases. The filters prevent
our tool from wrongly flagging 31 out of the 34 selected cases, while
3 cases which should have been flagged by the tool become false negatives.
The tool's reliability is improved with the small price paid of a
few false negatives being inserted. The filters thus prevent the tool
from flagging many more false positives. 

\subsection{Confounding Factors in Usage of Casts (RQ\protect\textsubscript{\textbf{3}})
\label{subsec:Data-Progression}}

\begin{figure*}[t]
\subfloat[Static Cast Assignment Cases.]%
{\includegraphics[width=1\columnwidth]{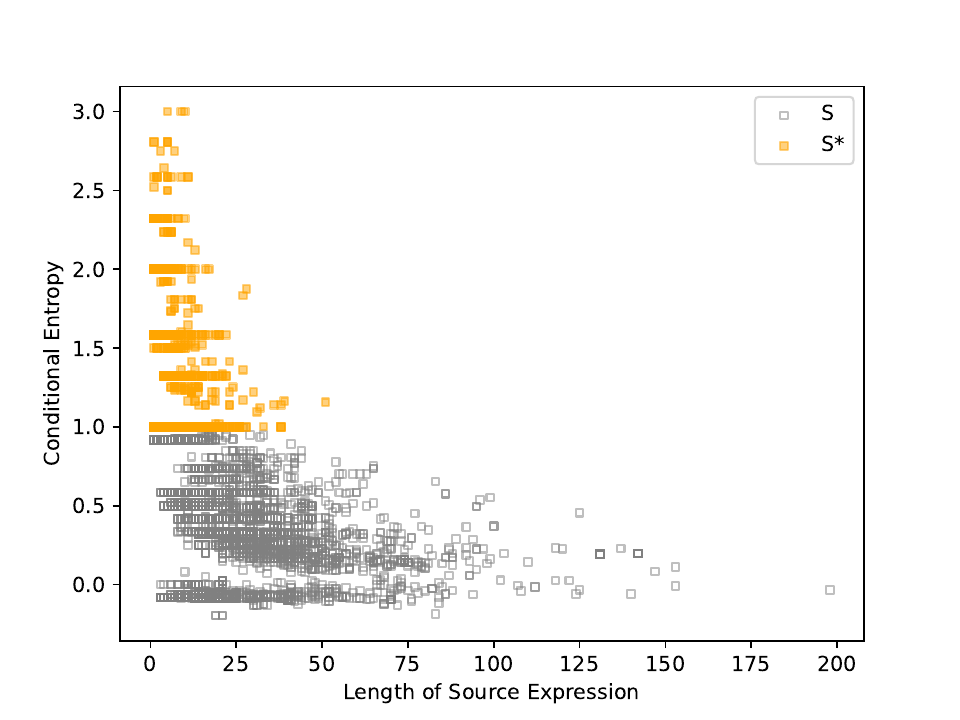}

\label{fig:a_rhs_ce_s}}%
\subfloat[Reinterpret Cast, Const Cast and Dynamic Cast Assignment Cases.]%
{\includegraphics[width=1\columnwidth]{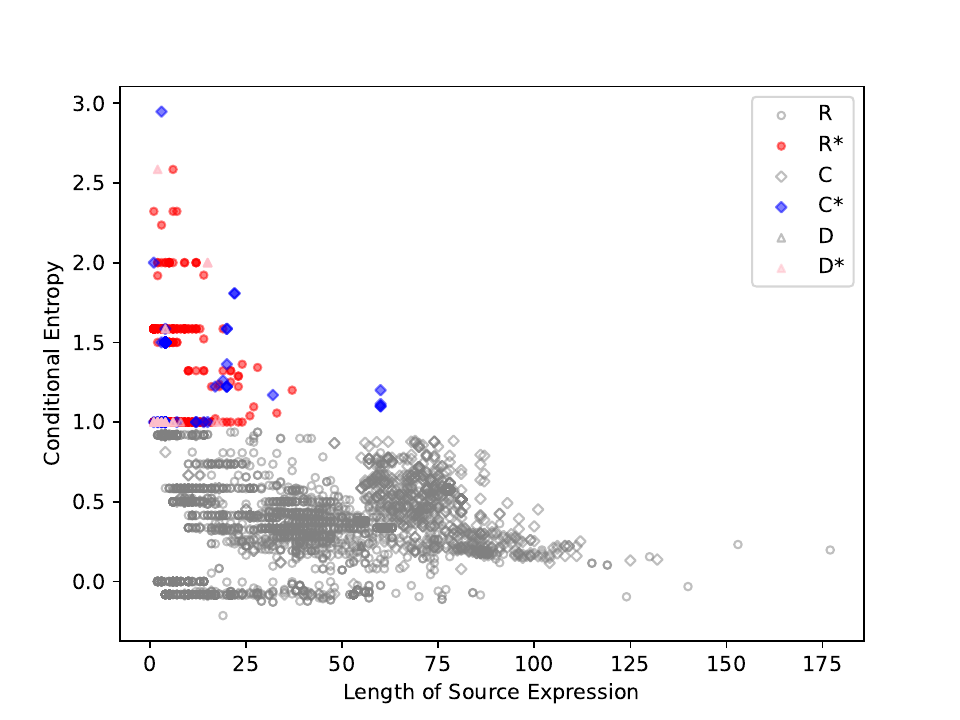}

\label{fig:a_rhs_ce_rest}}

\subfloat[Static Cast Function Call Cases.]%
{\includegraphics[width=1\columnwidth]{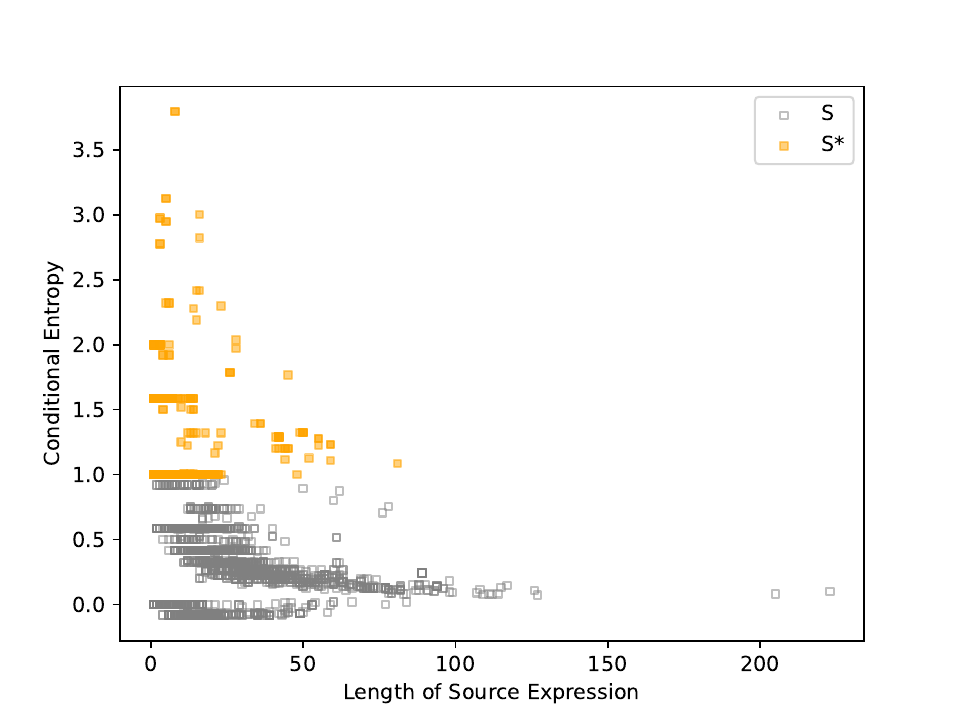}

\label{fig:c_rhs_ce_s}}%
\subfloat[Reinterpret Cast, Const Cast and Dynamic Cast Function Call Cases.]%
{\includegraphics[width=1\columnwidth]{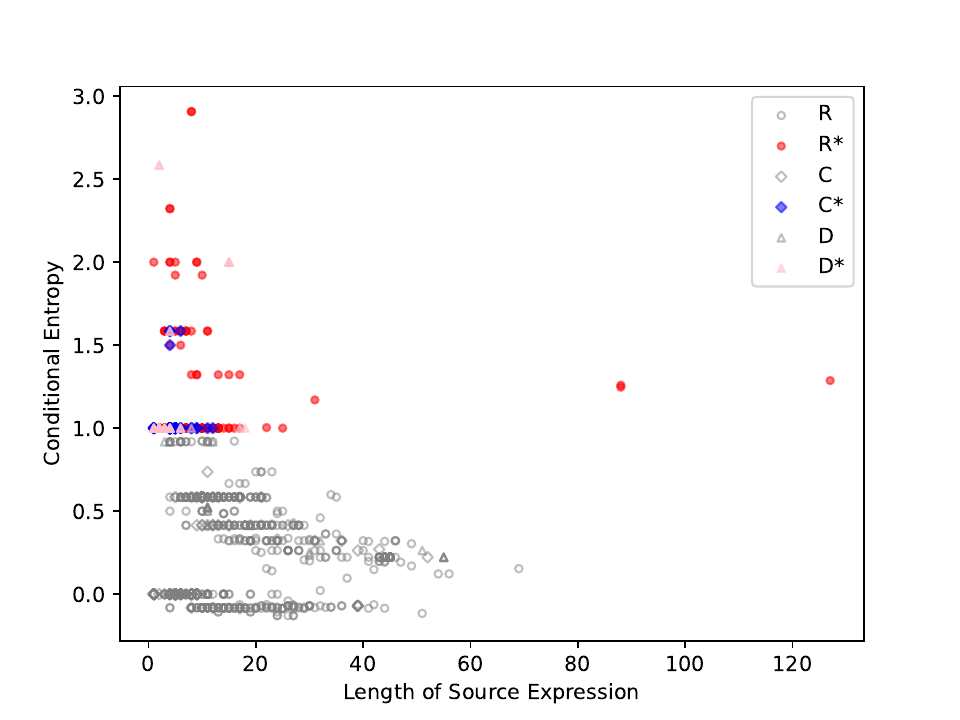}

\label{fig:c_rhs_ce_rest}}\caption{Type conversions represented by \emph{source} expression length and
conditional entropy. The star cases are the outliers.}
\end{figure*}

We associate the poor usage of casts with cases that present a higher
conditional entropy value. Thus, to answer \textbf{RQ}\textsubscript{\textbf{3}},
we inspect some long source expressions cases and we perform a correlation
analysis between source length and conditional entropy value. Poor
usage of casts may occur when developers have difficulty interpreting
the \emph{source} expression. This often happens for long and more
complex \emph{source} expressions. We used the number of characters
in the \emph{source} expression as a proxy of how complex a \emph{source}
is, and we investigated if casts with long \emph{source} expressions
are flagged by the tool. We chose characters over subtokens in the
\emph{source} to better capture the semantics in the identifiers.
For instance, \lstinline*relativeVelocity* and \lstinline*relV*
have the same amount of subtokens but \lstinline*relativeVelocity*
has more characters and is more descriptive.

\hyperref[fig:a_rhs_ce_s]{Figures} \ref{fig:a_rhs_ce_s}, \ref{fig:a_rhs_ce_rest},
\ref{fig:c_rhs_ce_s} and \ref{fig:c_rhs_ce_rest} show the conditional
entropy against the number of characters in the \emph{source}. The
highlighted cases are those with a higher conditional entropy value
than the threshold and are flagged by the tool. The \emph{source}
expression lengths range from 1 to 223 characters. Longer \emph{sources}
are a consequence of conditional blocks that the conditional operator
\lstinline*"?"*, or macro functions that generate long expressions
after compiler preprocessing. 

We performed a correlation analysis between the length of the source
and conditional entropy value. We computed the values of the following
two metrics: Pearson's correlation coefficient ($r$) and Spearman's
correlation coefficient ($r_{s}$). Pearson's correlation coefficient
measures if two variables are linearly related {[}\citeauthor{pearsonr}
\citeyear{pearsonr}{]}, while Spearman's correlation coefficient
measures the rank correlation of two variables {[}\citeauthor{spearmanr}
\citeyear{spearmanr}{]}. These coefficients can take values between
-1 and 1. If the value is 0, it indicates no correlation between the
variables. As the coefficient gets closer to either -1 or 1, the stronger
the correlation is between variables. In our case, $r$ has a value
of $-0.474$, while $r_{s}$ is $-0.587$ and both have $p<0.001$.
As the $r_{s}$ value is closer to -1 than $r$, we can notice that
the trend of the cases preserves a more monotonic correlation rather
than a linear correlation. The correlation coefficient values show
that the correlation between length of the \emph{source} and conditional
entropy is weak. 

As can be seen in \hyperref[fig:a_rhs_ce_s]{Figures} \ref{fig:a_rhs_ce_s},
\ref{fig:a_rhs_ce_rest}, \ref{fig:c_rhs_ce_s} and \ref{fig:c_rhs_ce_rest},
conditional entropy of the \emph{destination} somewhat decreases as
the \emph{source} length increases because longer identifiers hold
more information. This trend is also presented through the weak negative
correlation of the two variables. Surprisingly, the flagged cases
are largely limited to smaller \emph{source} expressions. A likely
interpretation for this is that the developers deliberately cross
type boundaries, using casts to change types. It is also possible
that the shorter \emph{sources} are not expressive enough and therefore,
might be misinterpreted by the developer. To better understand the
nature of developers' choices, we performed an analysis of hand-picked
named cast operations which is presented next.

\subsection{Qualitative Analysis (RQ\protect\textsubscript{\textbf{4}}) \label{subsec:Qualitative-analysis}}

To answer \textbf{RQ}\textsubscript{\textbf{4}}, we present a number
of interesting cases selected by our raters, followed by a summary
of how each operator was used. We grouped these cases into the following
categories: good implementations with precise names, poor implementations
with precise names, good implementations with imprecise names and
poor implementations with imprecise names. \\

\paragraph{{\small{}}\noindent {\small{}\textbf{Good Implementations with Precise Names}}}

The cast operations presented in this category are examples of good
implementations of the cast operators. The names of the source and
destination variables present a connection between them and the code
such that they are precise to the context. The cases have the following
conditional entropy values: Case 1 - 1.58, Case 2 - 2, Case 3 - 0.92,
Case 4 - 1.5 and Case 5 - 0.92. Even if some cases present a large
conditional entropy value, most of the time, those cases will not
be flagged due to the preprocessing filters. Cases 1 and 4 are detected
by the filter Casts from Literals by observing that the source is
a literal. Cases 2, 3, 4 and 5 are detected through the filter Descriptive
Variables and Type by noticing that the source identifier is part
of the destination identifier. 

\subsection*{Case 1 - {\large{}\lstinline*static_cast*} case}

\begin{lstlisting}%
[caption={The \lstinline*static_cast* operator is required from a portability
view to compare a pointer of type \lstinline*nullptr_t* with another
pointer of type \lstinline*Histogram*. This example presents a good
utilisation of \lstinline*static_cast* with precise names to perform
a safe pointer initialisation for \lstinline*atomic_histogram_pointer*.
This case has a conditional entropy of $1.58$.},float=h,label={lst:static_histogram}]
webrtc::metrics::Histogram* prev_pointer =  
  rtc::AtomicOps::CompareAndSwapPtr( &atomic_histogram_pointer,
  static_cast<webrtc::metrics::Histogram*> (nullptr),
  histogram_pointer);    

static T* CompareAndSwapPtr(T* volatile* ptr, T* old_value, T* new_value) 
{ return static_cast<T*>( ::InterlockedCompareExchangePointer(
        reinterpret_cast<PVOID volatile*>(ptr), old_value, new_value));
} 
\end{lstlisting}

An example of a \lstinline*static_cast* where the \emph{source }(\lstinline*nullptr*)
and \emph{destination }(\lstinline*old_value*)\emph{ }look different
is presented in \prettyref{lst:static_histogram}. This is an Assign
Value case with \emph{source} being a value and \emph{destination}
being specific. The Listing contains a call to \lstinline*CompareAndSwapPtr*
as well as the definition for the same. This method is actually called
from within a macro function definition, \lstinline*RTC_HISTOGRAM_COMMON_BLOCK*.
The purpose of this macro function is to add the information passed
to the \lstinline*histogram_pointer* safely. If the memory where
\lstinline*histogram_pointer* points is empty, then the pointer will
be changed to point to the new memory address. Otherwise, the code
from Lines 1-4 will ensure that it points to a \lstinline*nullptr*.

The \lstinline*static_cast* used on Line 3 in \prettyref{lst:static_histogram}
is passed as a parameter to the function \lstinline*CompareAndSwapPtr*.
The function call is part of a pointer declaration. The newly declared
pointer \lstinline*prev_pointer* will become the output of the method
\lstinline*CompareAndSwapPtr*. This function makes use of the API
\lstinline*Interlocked CompareExchangePointer* from \lstinline*Windows*
which is used to perform a pointer comparison and swap atomically.
The code has to clear \lstinline*atomic_histogram_pointer*. So, the
API call ultimately will compare the pointer with a \lstinline*nullptr*.
If those two pointers contain different values, then it will store
the value of \lstinline*nullptr* in the address of \lstinline*atomic_histogram_pointer*.
The \lstinline*static_cast* converts the \lstinline*nullptr* to
the type \lstinline*webrtc::metrics:Histogram** for consistency.

Since the code from \prettyref{lst:static_histogram} tries to validate
if \lstinline*atomic_histogram _pointer* is \lstinline*null*, it
is required to compare the pointer with a null pointer literal: \lstinline*nullptr*.
In order to compare two pointers, they need to be of the same type
and therefore, a \lstinline*static_cast* is used as it is the only
named cast operator which allows casts from \lstinline*nullptr* to
a different type. The \emph{destination} identifier to which the named
cast is bound is \lstinline*old_value*. While \lstinline*old_value*
looks different to \lstinline*nullptr* and that is why our information-theoretic
analysis identified it, the method \lstinline*CompareAndSwapPtr*
is likely designed to be generic and accepting of many different pointer
types. Therefore, this use of named cast is sound. Our tool does not
flag this case based on the preprocessing filters, since \lstinline*nullptr*
is a literal. Additionally, the \emph{source} and \emph{destination}
identifiers are specific to the context. Thus, we consider this case
as having precise names.

\subsection*{Case 2 - {\large{}\lstinline*dynamic_cast*} case}

\begin{lstlisting}%
[caption={In order to implement an exception handler, a \lstinline*dynamic_cast*
is necessary to prove a cast is valid. Generic names for the variables
are reasonable since the code is part of the Application Binary Interface
for C++. This case has a conditional entropy of $2$.},float=h,label={lst:dyn_buildtools}]
const __pointer_to_member_type_info* member_ptr_type =
    dynamic_cast<const __pointer_to_member_type_info*> (__pointee);
\end{lstlisting}

Since the sampled dataset had only one instance of \lstinline*dynamic_cast*,
we expanded our investigation to the entire dataset and analysed a
total of 11 cases. We present two cases of Check and Cast. The first
instance can be found in \prettyref{lst:dyn_buildtools}. It has been
extracted from \emph{private\_typeinfo.cpp} and it is part of \emph{libc++abi}
library. The use of the \lstinline*dynamic_cast* operator appears
in variable declarations in methods \lstinline*can_catch* and \lstinline*can_catch_nested*.
These methods are used for exception handling and report mismatches
during type conversions by checking if the result is null or not.
If it is, the methods return an exception. The \emph{source} variable,
in our example, has the identifier \lstinline*__pointee*, which is
of the type \lstinline*const __shim_type_info** . The \emph{destination}
variable is \lstinline*member_ptr_type*, which is a \lstinline*const*
pointer to \lstinline+ __pointer_to_member_type_info+, which itself
is derived from the class \lstinline*__pbase_type_info* a sub-class
of \lstinline*std::type_info* which contains information about types
for variables. The names in this cast are generic, however they are
still relatable to each other. Thus, we consider them to be precise.
\emph{libc++abi} implements the Application Binary Interface for \emph{C++}
and is expected to be generic to fit in with a wide spectrum of low-level
transactions between the application, libraries and the operating
system. The \lstinline*dynamic_cast* operator is used in this case
to check at runtime if the \emph{destination} variable can take the
\emph{source}'s type while keeping the natural language identifiers
as generic as possible. 

\subsection*{Case 3 - {\large{}\lstinline*dynamic_cast*} case}

\begin{lstlisting}%
[caption={An adequate example of \lstinline*dynamic_cast* performing a down-cast
conversion with meaningful names for \emph{source} and \emph{destination}.
This case has a conditional entropy of $0.92$.},float=h,label={lst:dyn_fmt}]
const DecimalFormat *decFmt = dynamic_cast<const DecimalFormat *>(&fmt);
\end{lstlisting}

The second example of Check and Cast is presented in \prettyref{lst:dyn_fmt}.
The snippet is from the file \emph{upluralrules}.\emph{cpp }in the
\emph{ICU} (International Components for Unicode) module. The \emph{source}
variable is \lstinline*fmt* with the type \lstinline+const class icu_64::NumberFormat*+
which captures the format of the expression. The \emph{destination}
variable is \lstinline*decFmt* and it has the type \lstinline+const class icu_64::DecimalFormat*+
. The \emph{destination}'s type class \lstinline*DecimalFormat* inherits
from \emph{source}'s type class \lstinline*NumberFormat* {[}\citeauthor{ICU_doc}
\citeyear{ICU_doc}{]} and this is an example of a down-cast operation
which is verified at runtime. If the checks fail and \lstinline*decFmt*
is \lstinline*NULL*, the method continues to check for other known
formats. The \emph{ICU} module handles a wide variety of data types.
Even for numerics, which is the focus of our example, there are several
different types that need checking: \lstinline*int32_t*, \lstinline*double*
and \lstinline*FixedDecimal*. Most of these values are only available
at runtime and therefore, the developers prefer to insert explicit
checks through the \lstinline*dynamic_cast* operator. The identifiers
in this case reflect the type specialisation that is happening through
the \lstinline*dynamic_cast* operator. This specialisation provides
us with precise names for the context. This is an example where type
conversions are used judiciously with clear objectives and the names
reflect the type conversion that is taking place. Further, the use
of \lstinline*dynamic_cast* operator makes the type conversion safe
at runtime. The preprocessing filters result in our tool not flagging
this case due to the \emph{source} identifier being a substring of
the\emph{ destination} identifier.\\

\subsection*{Case 4 - {\large{}\lstinline*const_cast*} case}

\begin{lstlisting}%
[caption={A fair example of how \lstinline*const_cast* is used to obtain a
non \lstinline*const* object from the \lstinline*const* pointer
\lstinline*this*. The generic identifiers for \emph{source} and \emph{destination}
indicate the code's purpose. This case has a conditional entropy of
$1.5$.},float=h,label={lst:const_nonConst_tznImpl}]
TimeZoneNamesImpl *nonConstThis = const_cast<TimeZoneNamesImpl *>(this);
\end{lstlisting}

There were only five cases of Modifiability in the sampled dataset.
Four cases belong to the library \emph{ICU} in two different files:
\emph{tznames\_impl.cpp} and \emph{tzfmt.cpp}. For these cases, the
\emph{source} identifiers are generic and partially different compared
to the \emph{destination} identifiers. \prettyref{lst:const_nonConst_tznImpl}
presents one of the four cases from the file \emph{tznames\_impl}.cpp.
The \emph{source} variable is the pointer \lstinline*this* which
is an instance of the class encapsulating the statement and has the
type \lstinline+const TimeZoneNamesImpl*+ . The \emph{destination}
variable is a pointer called \lstinline*nonConstThis* which does
not have the qualifier \lstinline*const* in its type. The chosen
identifiers for \emph{source} and \emph{destination} reinforces our
hypothesis that identifiers carry meaning. Here, the getters in the
encapsulating class need to maintain the integrity of the original
object. Thus, the desired values need to be extracted from a non \lstinline*const*
object derived from the pointer \lstinline*this* using a \lstinline*const_cast*
operator. This is an instance where explicit casting is being used
judiciously, clearly indicating its purpose through meaningful identifiers.
Our tool does not flag this case due to the preprocessing filters
recognising that the \emph{source} identifier is a substring of the\emph{
destination} identifier.

\subsection*{Case 5 - {\large{}\lstinline*reinterpret_cast*} case}

\begin{lstlisting}%
[caption={An example of \lstinline*reinterpret_cast* that is used in serialisation/deserialisation
to be able to iterate over an enumeration to reduce the amount of
code developers had to write. The identifiers used are precise and
the cast is well implemented. This case has a conditional entropy
of $0.92$.},float=h,label={lst:reint_dawn}]
auto memberBuffer = reinterpret_cast<DawnTextureFormat*> (*buffer);

for (size_t i = 0; i < memberLength; ++i) {
    memberBuffer[i] = record.colorFormats[i];
}
\end{lstlisting}

\prettyref{lst:reint_dawn} presents a case of Rapid Prototyping which
uses \lstinline*reinterpret_cast*. This snippet is from component
Dawn in file \emph{WireCmd\_autogen.cpp }and is one of 13 similar
cases. The file is generated from \emph{WireCmd}.\emph{cpp} using
the build system and contains serialisation and deserialisation functions.
The generated file is large with 14,000 lines of code and has a total
of 200 type conversions which have the same identifier for \emph{source}
variables and also for the \emph{destination} variables. The \emph{source}
identifier is the string \lstinline*buffer* and in most cases, it
is a pointer to a pointer for \lstinline+char+. There are cases when
the \emph{source} variables have additional type qualifiers such as
\lstinline*const volatile*. The \emph{destination} variable is \lstinline*memberBuffer*
and it is declared with the type \lstinline*auto*. We observed that
the \emph{destination} type varies from pointers to numeric types
such as \lstinline*unsigned long long* to pointers for structures
and enumerations. The casts are part of assignment expressions in
which the \lstinline*memberBuffer* is initialised with a part of
the \lstinline*buffer*. We consider the names to be precise because
the \emph{destination} identifier is more specific than the\emph{
source} and both names provide context to the cast. As with the previous
two cases, the \emph{source} identifier is a substring of the \emph{destination}
identifier and as such, the tool does not flag this case. 

The purpose of these casts is to serialise and deserialise a variety
of different structures for the component \emph{Dawn}. In other words,
the methods provide the functionality to convert objects to streams
of bytes and recreate the objects when needed. Since the universe
of types to be serialised is large, developers have relied on macros
to serialise/deserialise objects. The example selected in the \prettyref{lst:reint_dawn}
presents the \lstinline*buffer* which is converted in the type \lstinline*DawnTextureFormat*.
The target type is an enumeration. Similar to the example from \hyperref[subsec:An-example-of]{Section 2.2},
Lines 2-4 iterate over the enumeration. While the use of macros is
preferred for serialisation and deserialisation, given the massive
number of types that need to be serialised or deserialised, macros
provide little insight into the actual role of the casts. Nonetheless,
the generated file can be created from only 700 lines of code which
contain macros. The use of \lstinline*reinterpret_cast* in this case
is clearly beneficial from a software reuse point of view and leads
to a decrease in the amount of code. On the other hand, the named
cast operator is used to bypass the lack of an iterator for the enumeration
type, which if not done correctly, can be pernicious as \lstinline*reinterpret_cast*
comes with no semantic checks at all and as discussed above, enum
types may not be contiguous in the first place.\\

\paragraph{{\small{}}\noindent {\small{}\textbf{Poor Implementations with Precise Names}}}

The cast operations presented in this category are examples of poor
implementations of the cast operators, while the names are still precise
to the context. The cases have the following conditional entropy values:
Case 1 - 1.32, Case 2 - 2, Case 3 - 1.58 and Case 4 - 1.58. The tool
flagged all the cases due to a higher conditional entropy than the
threshold value of 1. The only exception is Case 4, which due to the
Casts from Literals filter, would be erroneously not flagged.

\subsection*{Case 1 - {\large{}\lstinline*static_cast*} case}

\begin{lstlisting}%
[caption={An example of \lstinline*static_cast* operator used in function \lstinline*ParseV4*
to parse an IPv4 address. The identifiers are precise, but the cast
was unnecessarily implemented. The \lstinline*ParseV4* function has
been refactored and the conversion has been replaced with the function
\lstinline+sscanf+ to parse the values. This case has a conditional
entropy of $1.32$.},float=h,label={lst:static_ipv4}]
address.bytes_[i++] = static_cast<uint8_t>(next_octet);
\end{lstlisting}

\prettyref{lst:static_ipv4} presents an Assign Value case of the
\lstinline*static_cast* operator in the component \emph{Base}, in
file \emph{ip\_address.cc} inside the method\emph{ }\lstinline*ParseV4*.
This method is used as part of the constructor for the class \lstinline+IPAddress+
to extract the IPv4 address from a string. The named cast operation
in \prettyref{lst:static_ipv4} is part of a variable assignment.
Although the \emph{source} and \emph{destination} identifiers are
selected because they look different, we need to understand how they
are used to assess whether a named cast is necessary here. We studied
how the \emph{source} and \emph{destination} identifiers are used
and found that the input string for \lstinline*ParseV4* is split
in octets in order to be parsed and added to the IPv4 address. The
\emph{source} identifier is \lstinline*next_octet* of type \lstinline*uint16_t*,
which represents one byte of the IPv4 address. The destination variable
is \lstinline*address.bytes_* where \lstinline*bytes_* is a member
of the class \lstinline*IPv4*. Specifically, it is an array of type
\lstinline*array<uint8_t, 16>*. The array has the length 16 since
\lstinline*IPAddress* can also have the IPv6 format. These identifiers
are self-explanatory within the code, thus we consider them precise.
The implementation of \lstinline*ParseV4* does not seem to be erroneous.
However, the use of the \lstinline*static_cast* operator is unnecessary
since the conversion from string to octets can be done using the built-in
type transformation type. Developers can use functions such as \lstinline+sscanf+
to read parts of the formatted string and return directly the desired
output. In fact, this is exactly what the developers did in later
versions of the implementation: the \lstinline*ParseV4* function
has now been refactored {[}\citeauthor{commit_ipadd} \citeyear{commit_ipadd}{]}
and updated to use \lstinline+sscanf+. \\

\subsection*{Case 2 - {\large{}\lstinline*reinterpret_cast*} case}

\begin{lstlisting}%
[caption={An example of \lstinline*reinterpret_cast* operators with precise
identifiers. These are used to allow functions to take pointer parameters
that can point to two different data structures. Thus, the function
returns a different result based on the variable \lstinline*flavor*.
It is not desirable to have pointers which point at different types
stuctures. This case has a conditional entropy of $2$.},float=h,label={lst:reint_mach}]
//check the total number of pages currently in use and pageable.
kern_return_t kr = host_statistics(host.get(), HOST_VM_INFO,
    reinterpret_cast<host_info_t>(&data), &count);

kern_return_t host_statistics(host_t host_priv, host_flavor_t 
    flavor, host_info_t host_info_out, 
    mach_msg_type_number_t *host_info_outCnt);

//retrieve the number of current physical processors
kern_return_t kr = host_info(mach_host.get(), HOST_BASIC_INFO,
    reinterpret_cast<host_info_t>(&hbi), &info_count);

kern_return_t host_info (host_t host, host_flavor_t flavor, 
    host_info_t host_info_out, 
    mach_msg_type_number_t *host_info_outCnt)
\end{lstlisting}

This case presents two similar cases that appear in different components.
The cases belong to Conversions Between Classes category and use the
\lstinline*reinterpret_cast* operator. Two different \emph{source}
identifiers (\lstinline*&data*, \lstinline*&hbi*) are bound to the
same \emph{destination} identifier (\lstinline*host_info_out*). \prettyref{lst:reint_mach}
contains the calls and the signature for the function \lstinline*host_statistics*
and \lstinline*host_info*. These method calls have been collected
from the files \emph{process\_metrics \_mac.cc} from \emph{Base} component
and \emph{audio\_low\_latency\_input\_mac.cc} from \emph{Media} component.
The functions \lstinline*host_statistics* and \lstinline*host_info*
are defined in the \emph{Mach} library which contains services and
primitives for the OS X kernel. 

The role of the functions \lstinline*host_statistics* and \lstinline*host_info*
is to retrieve host-specific information. The function \lstinline*host_statistics*
in Line 2 obtains information about virtual memory for a host. The
\lstinline*host_info* method in Line 10 retrieves basic information
about a host such as the number of current physical processors for
the host. Both methods return a variable \lstinline*kr* of type \lstinline*kern_return_t*.
This variable is an integer which maps to a list of generic errors.
If the method is successful, then \lstinline*kr* would have the value
\lstinline*0*. Otherwise, it would have a different value which represents
a specific error. Most of the methods from the \emph{Mach} library
follow the same coding conventions and they have a similar format
to these two functions.

The \emph{source} variable for the first case has the generic identifier
\lstinline*&data*. Its type is \lstinline*vm_statistics_data_t*
which is a pointer to the structure \lstinline*vm_statistics* and
contains statistics on the kernel's use of virtual memory. The \emph{source}
identifier for the cast from Line 10 is \lstinline*&hbi* which is
the acronym for its type, \lstinline*host_basic_info*. \lstinline*&hbi*
is the address of a structure \lstinline*host_basic_info* which is
used to present basic information about a host. The two casts from
\prettyref{lst:reint_mach} have the same \emph{destination} identifier:
\lstinline*host_info_out* with type \lstinline*host_info_t*. 

\lstinline*host_statistic* can hold two different types of structure:
\lstinline*vm_ statistics* for virtual memory information and \lstinline*host_load_info*
for host processor load information. The \lstinline*flavor* keeps
track of the type of statistics desired. In this way, the functions
will treat each \emph{destination} variable differently based on the
variable \lstinline*flavor*. Implementing the functions in this manner
allows them to perform different operations based on the parameters
passed. The \emph{destination} identifiers are identical since the
functions \lstinline*host_statistics* and \lstinline*host_info*
follow the same coding conventions and have a similar format. Unfortunately,
if the developer is not careful to pass the correct match between
the type and the \lstinline*flavor* as parameters to the functions,
it may lead to a crash. Despite the \emph{source} and \emph{destination}
identifiers being meaningful for the context, this is a case where
rigorously adhering to a coding convention can cause confusion during
development. \\

\subsection*{Case 3 - {\large{}\lstinline*reinterpret_cast*} case}

\begin{lstlisting}%
[caption={An example of \lstinline*reinterpret_cast* that allows deletion of
the indirection of the shared pointer. The case has precise identifier
names taking into account the context. Due to the complex design of
the \lstinline*BackingStore*, the code was refactored and the cast
was removed. This case has a conditional entropy of $1.58$.},float=h,label={lst:reint_delete}]
// The backing store deleter just deletes the indirection, which downrefs 
// the shared pointer. It will get collected normally.
void BackingStoreDeleter(... void* info) {   
	std::shared_ptr<i::BackingStore>* bs_indirection = reinterpret_cast<std::shared_ptr <i::BackingStore>*> (info);   
	...
	delete bs_indirection;
}
\end{lstlisting}

The code from \prettyref{lst:reint_delete} presents a Rapid Prototyping
case through the use of a \lstinline*reinterpret_cast* in Line 4.
The snippet is collected from component V8 in file \emph{api.cc}.
The \emph{source} variable is a \lstinline+void*+ pointer with the
identifier \lstinline*info*, while the \emph{destination} variable
is a shared pointer with the identifier \lstinline*bs_indirection*
of type \lstinline+std::shared_ptr<i::BackingStore>*+. To understand
this case, first, we need to understand what the type \lstinline*BackingStore*
is. In caching, a backing store is represented by the copy of a data
in the memory, more specific in our case, a copy to an \lstinline*ArrayBuffer*
{[}\citeauthor{backingData} \citeyear{backingData}{]}. The named
cast operator is used to retrieve the shared pointer for \lstinline*BackingStore*
data, which will be deleted later in the same function. The \lstinline*BackingStore*
pointer is a shared pointer that can be accessed from the V8 and the
Embedder components of Chromium and generates a lifetime management
problem when both components hold pointers to the backing store data.
The code complexity is increased since the components can resize the
shared memory or transfer ownership from one component to another.
The unsafe ownership model of \lstinline*BackingStore* is prone to
errors, such as memory leaks and access of the pointers after deleting
them, which has eventually led to various bugs {[}\citeauthor{backingData_bug1}
\citeyear{backingData_bug1}; \citeauthor{backingData_bug2} \citeyear{backingData_bug2}{]}. 

The problems have been solved by refactoring the ownership model and
making the \lstinline*BackingStore* to own the shared pointers {[}\citeauthor{backingData}
\citeyear{backingData}; \citeauthor{backingData_commit} \citeyear{backingData_commit}{]}.
The previous implementation required each component to delete its
shared pointer instance through the method \lstinline*BackingStoreDeleter*.
The new version of the \lstinline*BackingStore* class counts the
shared pointers references and if the count reaches zero, then the
\lstinline*BackingStore* will delete the pointer. The named cast
operation, along with the function \lstinline*BackingStoreDeleter*,
was removed in the new implementation {[}\citeauthor{backingData_commit}
\citeyear{backingData_commit}{]}. While the named cast operation
was not directly causing the bugs, we can definitely say that it added
complexity to the code by asking each component to delete its shared
pointer instance, and eventually the code led to bugs. Even though
the \emph{source} and \emph{destination} identifiers (\lstinline*info*
and \lstinline*bs_indirection*) are different, we can notice there
is a semantic relation between the identifiers and they are precise
for the context. \lstinline*info* refers to the data and \lstinline*bs_indirection*
refers to backing store pointer which is the copy of the data. If
a semantic perspective were to be considered, it is likely that this
case would not have been identified. \\

\subsection*{Case 4 - {\large{}\lstinline*reinterpret_cast*} case}

\begin{lstlisting}%
[caption={An example of \lstinline*reinterpret_cast* using precise identifier
names. This case was used to add a default value to the \lstinline+HashMap+
entries due to the lack of empty value behaviour. The code was refactored
by implementing empty value behaviour and the cast was removed. This
case has a conditional entropy of $1.58$.},float=h,label={lst:reint_removed}]
// Old implementation
#define F(name, str)
... 
  HashMap::Entry* entry =                                                   
	string_table_.InsertNew(name##_string_, name##_string_->Hash());   
  entry->value = reinterpret_cast<void*>(1); 

// New Implementation
#define F(name, str)
...
  string_table_.InsertNew(name##_string_, name##_string_->Hash());   
\end{lstlisting}

\prettyref{lst:reint_removed} presents two versions of a macro function
\emph{F} collected from the file \emph{ast-value-factory.cc} of component
AST. The first version contains an Assign Value case using the \lstinline*reinterpret_cast*
operator on Line 6. The \emph{source} expression is an integer literal
representing the value \lstinline*1*. The \emph{destination} variable
is a \lstinline+void*+ pointer with the identifier \lstinline*entry->value*
and it points to the value of an \lstinline+entry+ in a \lstinline+HashMap+.
The function \emph{F} is used in the initialisation of \lstinline+HashMap+
objects and each entry is initialised with value \lstinline*1*. The
\emph{destination} identifier is self-explanatory in the context,
and so we consider this case as having precise names. 

The second version of the macro function \emph{F,} which is a refactored
version \citep{refactored_hashmap}, does not contain the named cast
operation. With the lack of the named cast operation along with the
information from the commit, we can tell that the new implementation
of the \lstinline+HashMap+ supports objects with empty values without
causing any errors. The named cast operation in the first version
was a workaround, without a proper way of defining the behaviour if
the entries did not have values. This means that the code in the first
version was error-prone in the case of empty values. A proper implementation
shows that the named cast operation is not needed in the current case.
Due to the preprocessing filters detecting a literal, this case would
be erroneously not flagged.\\

\paragraph{{\small{}}\noindent {\small{}\textbf{Good Implementations with Imprecise Names}}}

The cast operation presented in this category is an example of good
implementation of the cast operator, while the names are imprecise.
The tool flags Case 1 because it has a conditional entropy value of
1.58. 

\subsection*{Case 1 - {\large{}\lstinline*const_cast*}}

\begin{lstlisting}%
[caption={An adequate example of how a \lstinline*const_cast* operator is used
to add the \lstinline*const* qualifier to a variable. The \emph{destination}
identifier could be renamed to be more meaningful. This case has a
conditional entropy of $1.58$.},float=h,label={lst:const_frames}]
size_t frame_count = base::debug::TraceStackFramePointers(
      const_cast<const void**>(frames), 
      max_entries, skip_frames);

size_t TraceStackFramePointers(const void** out_trace,
      size_t max_depth, size_t skip_initial)
\end{lstlisting}

\prettyref{lst:const_frames} presents a fair example of a Modifiability
case with Non Indicative identifiers. This example is taken from component
\emph{Base} and belongs to the method \emph{CaptureStackTrace} which
is used to collect frames in the execution stack. It is interesting
and complements the one discussed in \prettyref{lst:const_nonConst_tznImpl}
because the type qualifier \lstinline*const* is being added to a
value in this case. In this case, the type conversion is a parameter
for the function call \lstinline*TraceStackFramePointers*. The function
in Lines 1-3 returns the total number of the frames for the stack.
The \emph{source} identifier is \lstinline*frames* which has the
type \lstinline+void**+ and it represents the pointer to the stack
frames. Line 5 of \prettyref{lst:const_frames} shows the function
declaration. The \emph{destination} identifier is \emph{}\emph{\lstinline+out_trace+}
with the type \lstinline+const void**+. Being able to check the stack
is vital for debugging but at the same time, the stack should be protected
during debugging. The \lstinline*const_cast* is required in this
case to protect the stack frames from inadvertent manipulation while
the developer is inspecting the stack. Here, we see an instance where
the cast is necessary but the identifier for the \emph{destination}
is not descriptive enough. A more meaningful \emph{destination} identifier
would allow for greater clarity in the code. \\

\paragraph{{\small{}}\noindent {\small{}\textbf{Poor Implementations with Imprecise Names}}}

The cast operations presented in this category are examples of poor
implementations of the cast operators, while the names are also imprecise.
Case 1 and Case 2 have conditional entropy values of 2.32 and 1.58.
Since the values are higher than the threshold value, those cases
are flagged by the tool.

\subsection*{Case 1 - {\large{}\lstinline*static_cast*} case}

\begin{lstlisting}%
[caption={Example of how a \lstinline*static_cast* is used on primitive types.
The \emph{destination} variable is originally a \lstinline*void*
pointer and may potentially be misused if the developer is unaware
of the various types it can represent. The lack of precise \emph{destination}
identifier increases the chance of such misuse. This case has a conditional
entropy of $2.32$.},float=h,label={lst:static_surface}]
((unsigned int*)element)[0] = static_cast<unsigned int>(r);
((unsigned int*)element)[1] = static_cast<unsigned int>(g);
((unsigned int*)element)[2] = static_cast<unsigned int>(b);
((unsigned int*)element)[3] = static_cast<unsigned int>(a);
\end{lstlisting}

The code from \prettyref{lst:static_surface} presents a set of four
Rapid Prototyping cases through \lstinline*static_cast* conversions
collected from the component \emph{Swiftshader} from the file \emph{Surface}.cpp.
We identified the casts because the \emph{source} identifiers are
very short compared to the \emph{destination} identifiers. These casts
are inside a method \lstinline*write* which contains a \lstinline*switch*
statement that writes the colour values (RGBA format) to a data structure.
The \emph{source} identifiers are \lstinline*r*\emph{, }\lstinline*g*\emph{,
}\lstinline*b* and \lstinline*a* of type \lstinline*float*, which
represent the colours red, green and blue, and the opacity value (alpha).
The \emph{destination} identifiers, which are originally \lstinline*void*
pointers, have the generic name \lstinline*element* because it may
point to arbitrary data types. However, notice in \prettyref{lst:static_surface},
\lstinline*element* has been implicitly cast to point to an \lstinline*unsigned int*
to match the type for the desired\emph{ destination }type. Implicitly
casting \lstinline*void* pointers at the point of use can be confusing.
This could lead to the variable \lstinline*element* being treated
differently, assuming it has another type. Further to this, the \emph{destination}
identifier is not precise enough which further adds to the confusion.
We have found 45 similar conversions in the same \lstinline*switch*
statement. \\

\subsection*{Case 2 - {\large{}\lstinline*static_cast*} case}

\begin{lstlisting}%
[caption={Example of a \lstinline*static_cast* case with an imprecise \emph{destination
}identifier. This cast was part of complex code that led to inefficient
code. Under specific circumstances, the code crashed by retrieving
information from an empty pointer. This code was refactored and the
cast was no longer used. This case has a conditional entropy of $1.58$.},float=h,label={lst:static_path_rendering}]
template <typename T>
std::vector<Path *> GatherPaths(..., const void *paths
...
const auto *nameArray = static_cast<const T *>(paths);
\end{lstlisting}

Another Rapid Prototyping case using \lstinline*static_cast* is presented
in Line 4 in \prettyref{lst:static_path_rendering}, which belongs
to the file \emph{Context.cpp} from component \emph{libANGLE.} The
\emph{source} variable is a pointer of type \lstinline+const void*+
with the identifier \lstinline+paths+ and it represents a vector
of potential paths from the Render Tree. The \emph{destination} variable
is a pointer of type \lstinline+const auto*+ with the identifier
\lstinline+nameArray+. This conversion is required to allow the conversion
of the \lstinline+paths+ vector in a target template type. The template
type is used as an argument to the named cast operator in Line 4 and
it appears in the function template declaration on Lines 1-2 in \prettyref{lst:static_path_rendering}.
The role of the function \lstinline+GatherPaths+ is to iterate through
all the \lstinline+paths+ and returns their names. This case belongs
to a larger and more complex piece of code that has the functionality
to validate the command buffer at path rendering. The developers decided
to stop supporting this feature since this rendering method had a
worse performance compared to the other rendering methods {[}\citeauthor{nameArray_bug1}
\citeyear{nameArray_bug1}{]}. In addition, under specific circumstances
this functionality was trying to retrieve information from an empty
pointer which was leading to a crash {[}\citeauthor{nameArray_bug2}
\citeyear{nameArray_bug2}{]}. This example shows that a named cast
conversion can be used correctly, but it might also add complexity
to the code, leading to inefficient and error-prone code. In addition,
the \emph{destination} identifier is imprecise because it is generic
compared to the \emph{source} identifier. This further decreases the
code's clarity.

\paragraph{Usage of casts (RQ\protect\textsubscript{\textbf{4}})\label{subsec:cast_usage}}

In this work, we presented a summary of the findings from the named
cast operators study. We have identified: two cases of iteration over
enumeration types (\prettyref{fig:motiv} and \ref{lst:reint_dawn}),
two cases of poorly named variables (\prettyref{lst:static_surface}
and \ref{lst:reint_dawn}), two instances of anti-patterns that have
been refactored in later versions of the software so that the named
cast operators were no longer used (\prettyref{lst:static_ipv4} and
\ref{lst:reint_removed}), two cases that increased the complexity
of the code which led to poor quality code and bugs (\prettyref{lst:static_path_rendering}
and \ref{lst:reint_delete}), two cases that enabled a function to
change behaviour based on the types of the pointer (\prettyref{lst:reint_mach}),
and two good programming practices for protecting values stored in
variables (\prettyref{lst:const_nonConst_tznImpl} and \ref{lst:const_frames}). 

The operator \lstinline*static_cast* is the most versatile and most
widely used operator for explicit type conversions. In \prettyref{fig:motiv},
we discovered the use of the \lstinline*static_cast* to iterate over
an enumeration, which is an abuse of the enumeration type and an inefficient
implementation. \prettyref{lst:static_histogram} presents a good
use of \lstinline*static_cast*, demonstrating how it can be used
to provide safety during pointer initialisations. We also found examples
where named casts were used as a quick workaround. The case from \prettyref{lst:static_ipv4}
showed a cast which has been removed in recent versions. The case
from \prettyref{lst:static_surface} shows conversions between primitive
types, which in most cases is harmless.  However, the \emph{destination}
variable is a \lstinline*void* pointer which can point to many types
and lead to type confusions. The last case from \prettyref{lst:static_path_rendering}
shows a correct use of the \lstinline*static_cast* operator being
part of complex code that led to inefficient code and even to a bug.

The \lstinline*reinterpret_cast* operator is used mostly for pointer
to pointer conversions as it is the most permissive. \prettyref{lst:reint_mach}
presented two examples of conversions of two different pointer types
bound to a \emph{destination }which has the same name. Using the same
name to store data of different kinds is not desirable and we believe
the code can benefit from variable renaming. In \prettyref{lst:reint_dawn},
we presented an example of serialisation/deserialisation where the
developers have relied on \lstinline*reinterpret_cast* to be able
to deal with a diversity of objects. There is a strong software engineering
reason to do so as it is essential to keep the interface to the serialiser
and deserialiser generic to be able to deal with any data type. The
case from \prettyref{lst:reint_delete} shows another example where
complex code led to bugs. After the bugs were solved, the code was
refactored and the named cast was completely removed. Last case shows
the use of a \lstinline*reinterpret_cast* as a quick workaround to
not develop the behaviour for empty values case for entries of a HashMap.
This named cast operation was also removed in the recent versions.

\lstinline*dynamic_cast* operators are used infrequently. They are
used when the developer is unsure if a conversion is possible or not.
In this way, the runtime checks will confirm whether the casts are
valid. An example where it is mandatory to prove a cast is valid appears
in the implementation of an exception handler showed in \prettyref{lst:dyn_buildtools}.
Another essential use-case of \lstinline*dynamic_cast* operator is
for downcasts. The component ICU contains the most dynamic conversions
and they are used for downcasts. \hyperref[subsec:Related-Work]{Section 6}
discusses some solutions to avoid the expensive dynamic cast. However,
the question of why from all Chromium's components only ICU has implemented
its downcasts with \lstinline*dynamic_cast* remains unanswered.

The operator \lstinline*const_cast* is used for software engineering
reasons and security reasons. Even if this operator can introduce
undefined behaviour as presented in \hyperref[subsec:Context]{Section 2},
the analysed cases were adequately implemented. We have identified
two \lstinline*const_cast* usage patterns from the analysis. One
pattern appears when an object tries to access itself through the
pointer \lstinline*this* in a function declared with the qualifier
\lstinline*const*. The \lstinline*const* functions will make the
pointer \lstinline*this* also have the qualifier \lstinline*const*.
However, there are times when the \lstinline*const this* pointer
needs to be passed as a parameter to non-const functions. \prettyref{lst:const_nonConst_tznImpl}
shows an example where an explicit conversion was performed in a getter
to obtain information from an object. Another use-case appears when
some non-const variables need to be protected against modification
in specific methods. In order to do so, the \lstinline*const_cast*
is used to add the \lstinline*const* qualifier. \prettyref{lst:const_frames}
shows how a stack is passed as a parameter to a function after the
conversion. The motivation behind the use of some const type conversions
comes from the use of third party libraries.

\section{Threats to Validity\label{subsec:Limitations}}

\paragraph*{Internal threats}

The results of the manual investigation and the findings of the named
casts operators usages are influenced by the subjective experience
of the raters. We tried to minimise this bias by using three raters
with experience in C++. As mentioned in \hyperref[subsec:manual-evaluation]{Section 4.2},
each rater consulted the ISO C++ Standard {[}\citeauthor{iso_cpp}
\citeyear{iso_cpp}{]} to understand how the named cast operators
should be used, and only after this did the raters provide feedback
on the sample data. After each rater individually performed their
initial evaluation, they selected together the interesting cases presented
in \hyperref[subsec:Qualitative-analysis]{Section 4.5}.

\paragraph*{External threats}

Our tool is subject to analyse code where variable names are chosen
carelessly. In an ideal world, the natural language channel provides
enough context to understand the code's purpose. Our approach relies
on the connection between the identifiers to detect cast misuses and
the tool performs better if the identifiers are meaningful. In a scenario
where the names are chosen carelessly, our tool might identify fewer
cases of casts misuses, but it will identify more cases of imprecise
names. In many cases, cast misuse can be overshadowed by imprecise
naming. This is overcome by initially identifying imprecise naming,
essentially forming the first stage of a two stage refactoring - clarification
of intent followed by validation of intent. However, our tool will
also detect some false positives based on the nature of the approach.
Developers might decide in some cases that generic or different names
are appropriate for the \emph{source} and \emph{destination} identifiers.
In such cases, these casts would be flagged despite the identifiers
being meaningful to the code.

\selectlanguage{english}%

\section{Related Work\label{subsec:Related-Work}}

Research into type systems accelerated with Luca Cardelli's seminal
and accessible papers on type theory {[}\citeauthor{typefulPaper}
\citeyear{typefulPaper}; \citeauthor{understanding_types} \citeyear{understanding_types};
\citeauthor{types_data} \citeyear{types_data}{]}. He lucidly explained
how type systems could help us write better programs with fewer bugs.
Some of that research also discusses properties of types in object-oriented
programming. \citeauthor{explicit_casting_research} \citeyearpar{explicit_casting_research}
presented an analysis of the explicit type casts operators for C++
with details of each type of operator. \citeauthor{fastDynCastPaper}
\citeyearpar{fastDynCastPaper} proposed a method to implement dynamic
casts, which is an expensive operation, for systems where performance
is critical. \citeauthor{dynCastPaper} \citeyearpar{dynCastPaper}
have demonstrated the efficiency of the Gibbs and Stroustrup implementation
by using it as a baseline while also improving the performance by
a factor of two. 

\paragraph*{Type casting studies}

In term of the effects, there are a significant number of research
papers that present the study of the undefined behaviour introduced
by type conversions {[}\citeauthor{UB1} \citeyear{UB1}; \citeauthor{UB2}
\citeyear{UB2}; \citeauthor{UB3} \citeyear{UB3}{]}. Undefined behaviour
can have many causes and some of them are due to type conversions.
For instance, during the execution of a \lstinline*dynamic_cast*,
the program needs to check the pointer's type. This is done by the
dereferencing the pointer, and this case is undefined behaviour {[}\citeauthor{UBblog}
\citeyear{UBblog}; \citeauthor{UBblog2} \citeyear{UBblog2}{]}.
Compilers will capture some cases of undefined behaviour for which
they will generate warnings, but not all of them {[}\citeauthor{UB1}
\citeyear{UB1}{]}. For this reason, developers need tools and techniques
to verify their code.

\citeauthor{js_study} \citeyearpar{js_study} have done an empirical
study over the implicit casts for JavaScript. They proved that those
type conversions are in general harmless and developers use them correctly.
This can be translated as most of the times, implicit casts are safe
to use. However, there is contradicting evidence that unrestrained
named casts or explicit casts can have undesirable effects. Tools
have been researched and developed to detect such casts. \citeauthor{detectPaper_caver}
\citeyearpar{detectPaper_caver} present CAVER, which is a tool to
identify poor practices in casting and also discussed their security
implications. The tool analyses C++ code and focuses on the unsafe
uses of the \lstinline*static_cast* and \lstinline*dynamic_cast*.
This work has provided a good background to understand how named casts
can go wrong. Their tool's evaluation, much like ours, is performed
on the code from Chromium. \citeauthor{detectPaper_hextype} \citeyearpar{detectPaper_hextype}
provide another tool HexType that performs well at detecting badly
implemented casts. They have implemented HexType using low-overhead
data structures and compiler \foreignlanguage{british}{optimisations}
to minimise the required resources. \citeauthor{casting_java_explicit}
\citeyearpar{casting_java_explicit} provided an empirical study of
type conversions for Java. The target of their research is to discover
when and how developers use an explicit cast. This is done through
discovering and presenting 25 patterns of cast-usages from real-life
Java code. This paper is the closest to our work, but unlike us, it
does not use any signal from the natural language identifiers to detect
anti-patterns.

\paragraph*{Dual-Channel Research}

\citeauthor{knuth} \citeyearpar{knuth} proposed a paradigm shift
in programming, which is commonly known as Literate Programming, where
writing code to instruct a computer is secondary to presenting it
to human beings. In Literate Programming, each program contains its
explanation in natural language intermixed with sections of code.
Knuth presented the system WEB, which is a literate programming language
comprising of a document formatting language (TEX) and a programming
language (PASCAL). Literate programs contain a human-readable explanation
interspersed with code which is automatically picked up by the WEB
system to produce an executable. At the same time, WEB enables the
inclusion of powerful features such as pictures, equations, tables,
and others in the natural language part of Literate program. Thus,
the natural language information remains in harmony with the software
itself. 

Literate programming laid the foundation for novel research directions
in Software Engineering that drew upon advances in Natural Language
Processing. \citeauthor{naturalness} \citeyearpar{naturalness} proposed
the\emph{ naturalness hypothesis }for software which noted that large
programs can be repetitive and can be modeled with techniques that
capture repetition such as n-grams. They noted that code is analogous
to natural languages in the way it tends to repeat. Such repetitive
patterns can be harvested and interpreted as statistical properties
that can be used to develop better software engineering tools. They
used this observation to build a statistical language model over a
large corpus to improve code completion. An n-gram language model
was built using token sequences, which included natural language information
in the form of identifiers, from open source code. The model was used
in a plugin to complete code for Eclipse IDE which performed better
than the Eclipse's completion system at that time. 

Source code is normally written for it to run on a device. But, the
same code is also written for developers who maintain or improve the
application. Therefore, a large part of the code semantics is embedded
in the communication channels between developers i.e. the natural
language identifiers that are chosen and the comments that are written
in the code. Based on this insight,\emph{ }\citeauthor{dual_channel}
\citeyearpar{dual_channel} described two communication channels in
source code: the algorithmic channel (AL) and the natural language
channel (NL). The algorithmic channel comprises of all the instructions
written by the developers which will be executed by a computer. The
natural language channel, which consists of identifiers and comments,
provides information about the purpose of the code in a human-readable
format. The relation between the AL and NL channel can be utilised
to improve software analysis tools.

\citeauthor{flexeme} \citeyearpar{flexeme} have developed a tool
called HEDDLE to detect and separate tangled commits into atomic concerns.
HEDDLE generates a graph data structure that encodes different versions
of the program and annotates the data flow edges using the natural
language information from the source code. HEDDLE performs faster
and is more accurate in the detection of tangled commits than the
previous state-of-the-art. \citeauthor{posit} \citeyearpar{posit}
have also developed a technique called POSIT, which adapts NLP techniques
for tagging between code and natural language. POSIT can generate
more accurate tags for both source code tokens and natural language
words than the previous state-of-the-art. 

\paragraph*{Dual-channel Research On Extracting Meaning From Names}

Identifier names represent the majority of tokens from the source
code. \citeauthor{identifiers_code_quality} \citeyearpar{identifiers_code_quality}
have shown through an empirical study on Java applications that there
is a direct relation between the naming quality of identifiers and
source code quality. Thus, poor named identifiers show a lack of understanding
of the problem, which is translated into poor quality software. The
authors measured the quality of identifiers based on identifier naming
guidelines and subtokens comparison to Java and application specific
terms. Even if the subtokens' semantic meaning is ignored in the analysis,
this empirical study proves that the relation between the dual-channel
information is not entirely harvested and applied in software analysis
tools.

\citeauthor{LAP-def} \citeyearpar{LAP-def} defined the first \emph{Linguistic
Antipatterns (LAs)} which are a collection of recurring poor practices
in names of identifiers, documentation and code. LAs detect inconsistencies
based on semantic relations between names of variables, methods signatures
and the documentation. \citeauthor{LAP-impact} \citeyearpar{LAP-impact}
performed a study to discover how developers perceive LAs based on
the familiarity to the code. 69\% of the external participants (with
no familiarity to the code) and 51\% of internal participants (on
the development team for the code) rated LAs as poor development practices
that should be avoided. In addition, \citeauthor{LAP-cognitiveLoad}
\citeyearpar{LAP-cognitiveLoad} explored how developers' cognitive
load is influenced by the existence of LAs in code snippets. The authors
tracked the cognitive load using functional brain imaging and eye
movement tracking. The participants had to investigate code snippets
and identify semantic faults. The code snippets altered to contain
LAs significantly increased participants' cognitive load. Linguistic
Antipatterns, as proposed by Arnaoudova et al., motivate the development
of automatic tools such as the one we have developed in this paper.

\citeauthor{refinym} \citeyearpar{refinym} used dual-channel constraints
to mine conceptual types from identifiers and assignment flows between
them. Conceptual types are types that are latent in the program but
not explicitly declared by the developer. Generally, conceptual type
corresponds to the actual types, but there are cases where they can
be latent. For instance, password and username may have the same type,
\emph{string}, but their conceptual types are different. If a password,
which is generally a highly protected field, was declared the same
way as the username, it would lead to a vulnerability. 

\citeauthor{deepbugs} \citeyearpar{deepbugs} developed a learning
approach, called DeepBugs, for discovering bugs based on the semantic
meaning of the identifier names. This approach uses embeddings, a
vector representation for identifiers, which preserve the semantic
similarities between identifiers. The bug detection is treated as
a binary classification problem. DeepBugs approach trains a classifier
to distinguish correct code from incorrect code. The training data
consist of correct code and incorrect code generated by the authors.
The bug detectors use the embeddings from the training phase to discover
bugs. Three bug detectors were built based on this approach to discover
accidentally swapped function arguments, incorrect binary operators,
and incorrect operands in binary operations. The bug detectors have
a high accuracy between 89\% and 95\% to distinguish correct and incorrect
code. The bug detectors are also very efficient, taking less than
20 milliseconds to analyse a file. False positives are inevitable
in static analysis tools; however, the bug detectors have a 68\% true
positive rate. 

Another approach that makes use of the semantic meaning of the identifier
names is presented by \citeauthor{context2name} \citeyearpar{context2name}
and it is called Context2Name. JavaScript code is usually deployed
in a minified version in which the identifiers are replaced with short
and random names. Context2Name is a deep learning-based technique
that predicts identifier names for variables that have a minified
name. This technique generates context vectors for each identifier
by inspecting five tokens before and after the identifier's occurrence.
The context vectors are then summarised in embeddings. Those embeddings
are used by a recurrent neural network to predict natural names for
the minified variables. Context2Name predicts correct identifiers
with a 47.5\% accuracy of all minified names and it predicts 5.3\%
additional identifiers missed by the state-of-art tools. 

The improvements made by the dual-channel research shows how much
potential the dual-channel information presents for software analysis.
Our study uses similar approaches to the work from dual-channel research,
but on a different problem. Hints of the developer's intent have been
extracted from natural language information to guide the detection
of anti-patterns of named casts.\selectlanguage{british}%

\section{Conclusion\label{subsec:Conclusion}}

Identifiers can add insights into program semantics and can be used
for sanity checking cast operations where the developers cross type
boundaries. Our work provides insights into how developers use named
casts and can be used to prioritise refactoring for named cast operators.
The approaches presented in this work are lightweight and can be easily
used during development, as an IDE plugin during development or either
at the review stage. While our tool is for any C++ project, it depends
on descriptive names to make decisions and therefore, can potentially
perform better for projects with well-established coding guidelines.
These coding guidelines need to contain precise details for how the
identifiers should be named and regular code reviews should ensure
that developers are following these guidelines. 

The evaluation results and the in-depth analysis demonstrate the potential
of the dual channel approach for program analysis. Despite compelling
results, there is still room for improvement in our approach. We did
not find value in directly applying techniques from Natural Language
Processing for analysing sub-tokens in identifiers due to the domain-specific
nature of our corpus. For this reason, future work on our tool could
involve combining domain-specific knowledge with dual channel analysis
to improve the reliability of our tool. Despite this, our work also
provides a strong foundation to help richer forms of static analysis
scale by using novel program representations derived from the natural
language channel in software.

\subparagraph*{Acknowledgments}

This research was partly funded by EPSRC grant number EP/W015927/1.

\bibliographystyle{ACM-Reference-Format}
\bibliography{112_Users_costin_Desktop_phd_year3_writing_cast_paper-named-casts_explicit_cast_paper_lyx_ref}

\end{document}